\documentclass[aps,prb,article,twocolumn,preprintnumbers,amsmath,amssymb,superscriptaddress]{revtex4}

\date{\today}
\usepackage{epsfig}
\usepackage{subfigure}
\usepackage{graphicx}
\usepackage{dcolumn}
\usepackage{bm}
\usepackage[colorlinks,linkcolor=blue,hyperindex,CJKbookmarks]{hyperref}
\usepackage{float}
\usepackage{hyperref}
\usepackage{comment}
\usepackage{color}
\newcommand{\Ham}   {{\mathcal{H}}}

\newcommand{\rbf}      {\textbf{r}}
\newcommand{\Rbf}      {\textbf{R}}
\newcommand{\Schrdg} {{Schr\"{o}dinger}}

\begin{document}

\title{\emph{Ab Initio} Exact Diagonalization Simulation of the Nagaoka Transition in Quantum Dots}
\author{Yao Wang}
\email{yaowang@g.harvard.edu}
 \affiliation{Department of Physics, Harvard University, Cambridge, Massachusetts 02138, USA}
 \author{Juan Pablo Dehollain}
\affiliation{QuTech and Kavli Institute of Nanoscience, TU Delft, 2600 GA Delft, The Netherlands}
\affiliation{School of Mathematical and Physical Sciences, University of Technology Sydney, Ultimo NSW 2007, Australia}
 \author{Fang Liu}
 \affiliation{Department of Chemical Engineering, Massachusetts Institute of Technology, Cambridge, Massachusetts 02138, USA}
\author{Uditendu Mukhopadhyay} 
\affiliation{QuTech and Kavli Institute of Nanoscience, TU Delft, 2600 GA Delft, The Netherlands}
\author{Mark S. Rudner}
\affiliation{Center for Quantum Devices and Niels Bohr International Academy, Niels Bohr Institute, University of Copenhagen, 2100 Copenhagen, Denmark}
\author{Lieven M. K. Vandersypen}
\affiliation{QuTech and Kavli Institute of Nanoscience, TU Delft, 2600 GA Delft, The Netherlands}
\author{Eugene Demler}
\affiliation{Department of Physics, Harvard University, Cambridge, Massachusetts 02138, USA}
\date{\today}
\begin{abstract}
Recent progress of quantum simulators provides insight into the fundamental problems of strongly correlated systems. To adequately assess the accuracy of these simulators, the precise modeling of the many-body physics, with accurate model parameters, is crucially important. In this paper, we employed an \emph{ab initio} exact diagonalization framework to compute the correlated physics of a few electrons in artificial potentials. We apply this approach to a quantum-dot system and study the magnetism of the correlated electrons, obtaining good agreement with recent experimental measurements in a plaquette. Through control of dot potentials and separation, including geometric manipulation of tunneling, we examine the Nagaoka transition and determine the robustness of the ferromagnetic state. While the Nagaoka theorem considers only a single-band Hubbard model, in this work we perform extensive \emph{ab initio} calculations that include realistic multi-orbital conditions in which the level splitting is smaller than the interactions. This simulation complements the experiments and provides insight into the formation of ferromagnetism in correlated systems. More generally, our calculation sets the stage for further theoretical analysis of analog quantum simulators at a quantitative level.
\end{abstract}

\maketitle
\section{Introduction}
Strong correlations are at the heart of many important phenomena in condensed matter systems, including unconventional superconductivity\cite{bednorz1986possible}, quantum magnetism\cite{fazekas1999lecture} and fractional quantum Hall states\cite{stormer1983quantized}. These phenomena have a wide range of applications in material design, energy science, and quantum information\cite{dagotto1994correlated}. The complexity of strongly correlated many-body systems does not allow us to apply traditional theoretical approaches based on perturbation theory, and requires using hard-core numerical techniques, including exact diagonalization\cite{lin1990exact}, quantum Monte Carlo\cite{foulkes2001quantum}, density-matrix renormalization group\cite{schollwock2005density}, \emph{etc}. However, these numerical techniques are limited to restricted conditions such as small size, high temperature, and low dimension. The pursuit for understanding strongly correlated systems in materials motivates new approaches that can overcome these restrictions.

In addition to conventional numerical techniques, analog quantum simulators offer a distinct solution. Specifically, cold-atom simulators in optical lattices have achieved great success in simulating interacting bosonic systems\cite{bloch2008many, lewenstein2012ultracold, gross2017quantum} and have recently begun exploring fermionic systems\cite{parsons2015site, greif2016site, parsons2016site, boll2016spin, brown2019bad,nichols2019spin}. Taking advantage of electrons as charged particles, solid-state quantum-dot simulators naturally incorporate the Coulomb interactions and provide an alternative for mimicking electronic many-body states in molecules and solids\cite{van2002electron, hanson2007spins, nielsen2007nanoscale, oguri2007kondo, thalineau2012few, seo2013charge, hensgens2017quantum, mukhopadhyay20182}. With the relatively easy accessibility of high orbitals and low temperatures, the quantum-dot simulators are promising to simulate a realistic system. Despite the experimental progress with these platforms for quantum simulation, the interpretation of the underlying physics is still at the stage of minimal models with estimated parameters\cite{hilker2017revealing, mazurenko2017cold, barthelemy2013quantum}. This limits the quantitative analysis of fine details of experiments and hinders extensions to more complicated models. 

A solution to this problem might be readily available, if we turn to the fields of chemistry and material science, where atomic-basis-based \emph{ab initio} approaches have been well developed. The spirit of these approaches is the unbiased evaluation of all physical parameters from a given set of atomic ingredients. In the past half a century, \emph{ab initio} calculations have made great progress towards describing systems with increasing complexity. With the help of the Gaussian basis\cite{binkley1980self,ditchfield1971self,frisch1984self,hariharan1973influence,hehre1969self}, the computational cost has been largely reduced, making the simulation of large molecules possible. In addition to the basic Hartree-Fock method\cite{hartree1928wave,hartree1935self,fock1930naherungsmethode,fock1930selfconsistent,slater1930note}, many advanced post-Hartree-Fock wavefunction-based methods (coupled cluster\cite{shavitt2009many}, configuration interactions\cite{sherrill1999configuration,pople1987CI}, \emph{etc.}) and multi-reference methods\cite{hegarty1979application, eade1981direct, yamamoto1996direct, andersson1990second, andersson1992second} have been invented.  More recently, advanced computer architectures including graphical processing units (GPUs) have been widely exploited by quantum chemistry simulations, pushing the scale of calculation to even larger systems\cite{ufimtsev2008graphical,ufimtsev2008quantum, deprince2011coupled, hohenstein2015atomic, song2018reduced}. Though successful in chemistry, existing software packages are not compatible with quantum simulators: the state-of-the-art quantum chemistry calculations are based on existing atomic wavefunction bases; however, the tunability of quantum simulators requires the wavefunction basis being the eigenstates of given, arbitrary potential landscapes, which are obtained numerically during the calculation.

In small quantum-dot systems, initial progress has been made using fixed wavefunction bases. Early studies focused on the physics in a single parabolic quantum dot, whose basis wavefunctions are Fock-Darwin states -- a subset of Gaussian wavefunctions. With these analytical wavefunctions as bases, the many-body Hamiltonian can be easily computed and the ground-state solution can be obtained using exact diagonalization\cite{maksym1990quantum,merkt1991energy,pfannkuche1993comparison,hawrylak1993magnetoluminescence,yang1993addition,hawrylak1993single,palacios1994capacitance,wojs1995negatively,oaknin1995low,wojs1996charging,maksym1996eckardt,hawrylak1996magnetoexcitons,wojs1997theory,wojs1997spectral,eto1997electronic,maksym1998quantum,reimann2000formation,mikhailov2002quantum} and quantum Monte Carlo\cite{bolton1996fixed,harju1999many,harju1999wave,harju2002wigner,siljamaki2002various}. The simplified treatment was also extended to double-dot systems\cite{burkard1999coupled, li2009artificial,nielsen2012many,stepanenko2012singlet, yannouleas2016ultracold, brandt2017bottom}. Since the parabolic potentials cannot describe the ``crystal-field'' corrections -- the impact of neighboring dot potentials on the single-particle wavefunctions and site energies -- recent work has considered more realistic Gaussian potentials. In this case, density functional theory (DFT) and wavefunction based methods such as configuration interaction have been attempted, using a numerical wavefunction basis beyond the Fock-Darwin states\cite{abolfath2006real,stopa2006electronic,abolfath2006theory}. However, these are the largest quantum-dot systems that have been subject to \emph{ab initio} attempts. In trying to find a compromise between model accuracy and computational complexity, simulations of larger systems have been restricted to simpler toy models like the Hubbard and extended-Hubbard models\cite{stafford1994collective,stafford1997coherent,stafford1998coherent,kotlyar1998addition,kotlyar1998correlated}. 

A recent experiment~\cite{dehollain2019nagaoka} showcased some of the power of quantum dot based simulators for studying quantum magnetism, by using a 2$\times$2 plaquette to investigate Nagaoka magnetism
-- magnetism induced by a single hole in a half-filled correlated electronic system. This phenomenon has been difficult to realize experimentally, in great part because of the correlated nature of the electronic system required to observe the physics of Nagaoka ferromagnetism\cite{nagaoka1966ferromagnetism}. The success of the experiment in Ref.~\onlinecite{dehollain2019nagaoka} relied on pushing the limits of the maximum achievable interaction strengths, as well as the minimum measurable energy gaps. The observed energy gap crucial for Nagaoka ferromagnetism is of the order of a few $\mu$eV in such a system, three orders of magnitude smaller than the level spacing between orbitals and the ground-state Coulomb interaction. 

Given that Nagaoka ferromagnetism was proven in a single-band finite system, it is not obvious that this phenomenon should persist when the level spacing among different orbitals is well below the interaction scales, as is the case in the quantum-dot experiment by Dehollain \emph{et al}~\cite{dehollain2019nagaoka}. Thus, these system conditions require a precise numerical many-body approach in order to validate the experimental observations. Moreover, to reflect the tunability of quantum dots comparable with realistic experiments, the modeling with \emph{ab initio} inputs is also necessary. 

For both of these purposes, we hereby introduce an \emph{ab initio} exact diagonalization framework to describe artificial quantum simulator systems consisting of multiple quantum dots. By calculating the wavefunctions in a given potential well and evaluating the one-center and two-center integrals, we construct the tight-binding Hamiltonian of the many-body system consisting of multiple interacting quantum dots. This calculation predicts the single-particle energies, along with various interaction energies, which are quantitatively consistent with experiments\cite{dehollain2019nagaoka}. Additionally, we applied the calculation on a plaquette system, reproducing the experimental conditions that led to the observation of the Nagaoka ferromagnetic ground state. The model again shows good agreement with the experimentally observed energy gaps, as well as with the observed robustness of the ferromagnetic state performed in the experiment\cite{dehollain2019nagaoka}.

The description of this model and calculation will gradually increase in complexity. In Sec.~\ref{sec:singleWell}, we first explain the single-well wavefunction basis and the numerical implementation that automatically generates the basis based on a given potential. After that, we present the derivation and implementation of many-body Hamiltonians in multiple quantum wells in Sec.~\ref{sec:manybodyHam}. By adjusting the model to represent a four-well system, in Sec.~\ref{sec:calculations} we then explore the quantum magnetism and especially the Nagaoka transition using the \emph{ab initio} exact diagonalization approach. Finally, we conclude and discuss the future directions of our approach in Sec.~\ref{sec:conclusion}.

\section{Single Electron in a Single-Quantum Well}\label{sec:singleWell}
\begin{figure}[!t]
\begin{center}
\includegraphics[width=\columnwidth]{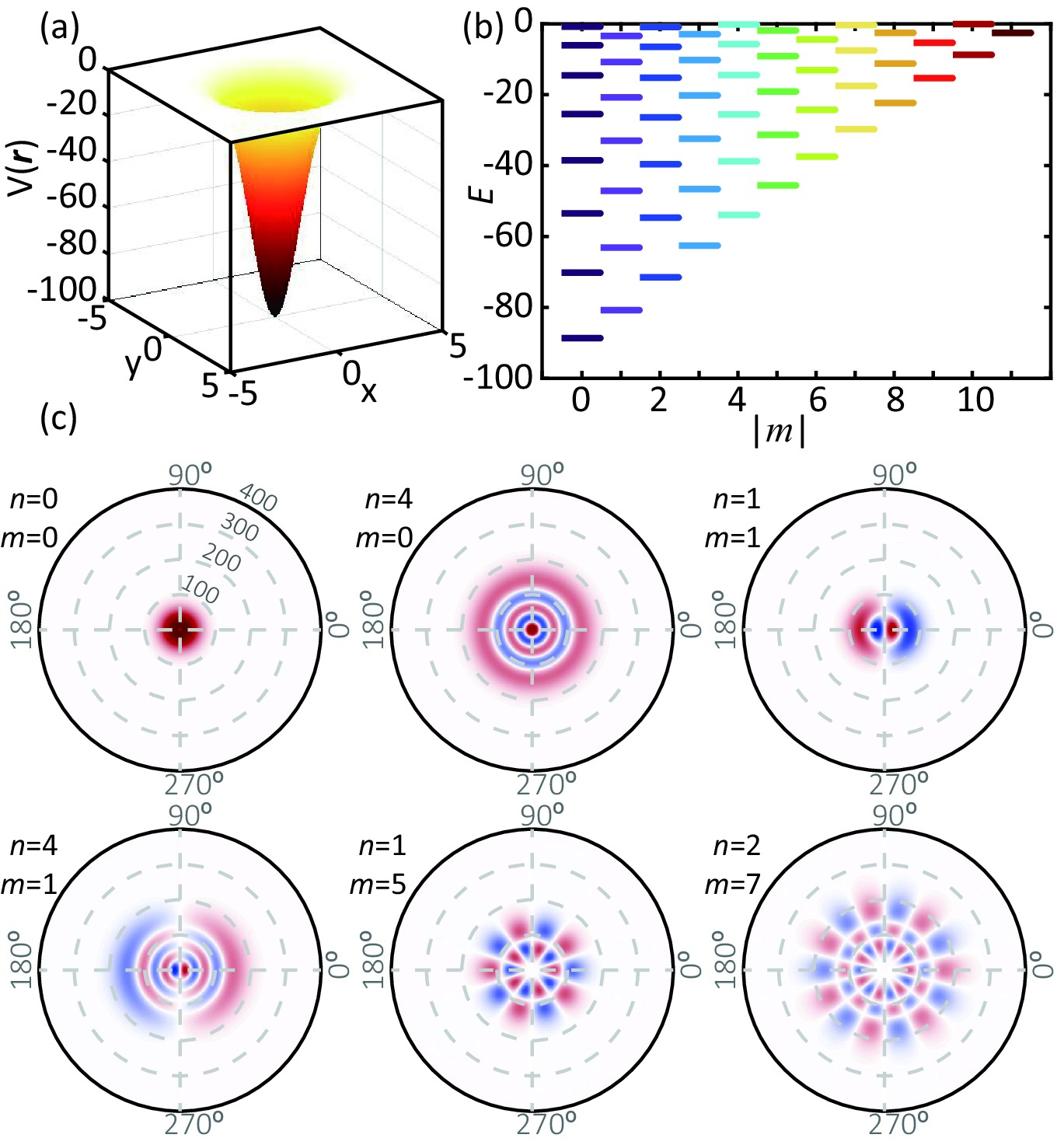}
\caption{\label{fig:1} Solution of single-well wavefunctions for $V_0 = 100$ and $\Sigma =1$. (a) The Gaussian quantum well in two dimensions. (b) Eigen-energy solutions for all bound states in the quantum well of (a), with the colors denoting different angular quantum numbers. (c) Sample eigenstate wavefunctions for $(n,m)=(0,0)$, (4,0), (1,1), (4, 1), (1,5), and (2,7), respectively.
}
\end{center}
\end{figure}
To simulate the electrons trapped in a finite-width potential well, we consider a confining central potential with rotational symmetry. This confining potential mimics the combined impact of electrodes surrounding the quantum dot\cite{potentialshape}. Though a generic potential landscape, obtained by solving the Poisson equation, can be employed as an input in the calculation, we use the Gaussian potential $V(\rbf) = -V_0 e^{-{|\rbf|^2}/{2\sigma}}$ in this paper [see Fig.~\ref{fig:1}(a)], as a typical description of the finite-size quantum dot\cite{abolfath2006theory,abolfath2006real}. Here $\rbf = r\cos\phi\, \mathbf{e}_x + r\sin\phi\, \mathbf{e}_y$ is the spatial coordinate with respect to the center of quantum well. In contrast to an atomic potential, the quantum well has finite potential energy with no singularity; unlike the parabolic potential, the Gaussian potential has a finite width and finite number of bound states. The single-electron static \Schrdg\ equation is
\begin{equation}
\left[-\frac{\hbar^2}{2m_e^\star}\! \left(\!\frac{\partial^2}{\partial r^2} + \frac1{r}\frac{\partial}{\partial r} + \frac1{r^2}\frac{\partial^2}{\partial \phi^2}\!\right)\! +\! V(r)\right]\! \psi(\rbf) = E\psi(\rbf)\,,
\end{equation}
where $m_e^\star$ is the effective mass of electron in the two-dimensional electronic gas (2DEG). The equation can be simplified by separation of variables
\begin{equation}
    \psi(\rbf) = \frac{\chi(r)}{\sqrt{2\pi r}} e^{im\phi} = \frac{\chi(r)}{\sqrt{r}}\varphi(\phi),
\end{equation}
where the $\chi(r)$ and $\varphi(\phi)$ are the radius and angular wavefunctions. Denoting the radial quantum number as $n$ and angular quantum number as $m$, the set of $\{\chi_n(r)\}$ satisfies the normalization condition
\begin{equation}
    \int_0^\infty \chi_n(r)^* \chi_{n^\prime}(r) dr = \delta_{nn^\prime}\,.
\end{equation}
Then we obtain the radial differential equation
\begin{equation}
    -\frac{\hbar^2}{2m_e^\star}\left[\frac{d^2\chi}{dr^2} - \frac{m^2-1/4}{r^2} \chi\right] + V(r)\chi = E\chi\,.
\end{equation}
It can be numerically solved using the finite difference approximation.
Choosing the angular part being real for numerical convenience, we define the \textit{single-well} wavefunction as
\begin{equation}\label{eq:singleWellBasis}
    \psi_{nm}(\rbf) = \left\{\begin{array}{ll}
        \frac{\chi_n(r)}{\sqrt{2\pi r}}, & m=0 \\
        \frac{\chi_n(r)}{\sqrt{\pi r}} \cos(m\phi), & m>0\\
        \frac{\chi_n(r)}{\sqrt{\pi r}} \sin(m\phi), & m<0.
    \end{array}\right.
\end{equation}
These eigenstate wavefunctions define the 2D orbital $(n,m)$ quantum numbers, while the spin component will be introduced later. We label the single-well single-electron eigen-energy as $\varepsilon_{nm}$. As shown in Fig.~\ref{fig:1}(b), the energy levels are well separated near the ground state, but become denser at higher energies. This is typical in a finite potential well. Unlike a parabolic potential, there are finite number of bound states (denoted as $N_{\rm orbital}$) in a finite well.

The wavefunctions of the eigenstates also become more extended with the increase of energy, or equivalently quantum numbers. While $m$ determines the angular distribution of a wavefunction, $n$ gives the number of nodes along the radius. Figure~\ref{fig:1}(c) shows examples of a few eigenstate wavefunctions. The ground state $(n,m)=(0,0)$ is restricted to the center of the potential well with a Gaussian-like shape, while the high-energy states such as $(n,m)=(2,7)$ spread three times wider.

Different from 3D systems, the eigenstates of a 2D potential well have two-fold orbital degeneracy for all $|m|>0$ (\emph{i.e.},~$p$, $d$, $f$ orbitals in atomic notation). This degeneracy is maintained in a $C_4$ symmetric system. This rotational-symmetric shape of the potential well is a theoretical simplification. In reality, the confining potential is not perfectly symmetric and can deviate from the solution in Fig.~\ref{fig:1}, resulting in the level splitting of the degenerate states\cite{van2005spin}. However, as we will show in Sec.~\ref{sec:calculations}, the ideal model gives an adequate estimation of the experimentally measurable parameters, both qualitatively and quantitatively. This result indicates that single-well wavefunctions obtained from the rotational-symmetric potential well also form a good basis to expand local electronic states.

\section{Many-Body Model}\label{sec:manybodyHam}
With multiple potential wells, the general Hamiltonian for a many-body system among $N_{\rm well}$ wells is 
\begin{equation}\label{eq:nonTighBindingHamiltonian}
\Ham = \sum_a^{N_e} \left[ -\frac{\hbar^2}{2m_e^\star} \nabla^2_a + \sum_i^{N_{\rm well}} V(\rbf_a - \textbf{R}_i)\right] + \sum_{a\neq b} \frac{e^2}{4\pi\epsilon|\rbf_a -\rbf_b|}\,,
\end{equation}
where the sum over $a$ and $b$ traverses the $N_e$ electrons, while the sum over $i$ traverses different potential wells. The first term is a sum with respect to each electron, which can be treated by separation of variables. Different from a chemistry problem, here the electrostatic potential $V(\rbf)$ is given by the electrodes and there is no need to introduce the Born-Oppenheimer approximation. 

Following the linear combination of atomic orbitals (LCAO) approach in the electronic structure theory\cite{szabo2012modern}, we construct the basis using a superposition of the single-well wavefunctions
\begin{equation}\label{eq:orthTransform}
\tilde{\psi}_{\mu\sigma}(\rbf) = \sum_{\nu} X_{\nu\mu}\psi_{\nu\sigma}(\rbf)\,.
\end{equation}
For simplicity in notation, we collapse the coordinate and orbital indices as $\mu = (i,\alpha)$, and denote $\psi_{\mu\sigma}(\rbf) = \psi_\alpha(\rbf -\mathbf{R}_i)s(\sigma)$. The $s(\sigma)$ denotes the spin wavefunction which does not mix in the hybridization.
Since the single-well wavefunctions are truncated at a relatively high level, this linear combination does not span a complete spatial basis, but is enough for the ground-state calculation when the number of tracked orbitals is much larger than the number of occupied orbitals.
With the presence of multiple  wells, the single-well wavefunctions are no longer orthogonal. An orthonormalization should be applied in order to simplify the many-body \Schrdg\ equation. The overlap matrix among different single-well wavefunction basis is 
\begin{equation}
\int d\rbf^3 \psi_{\mu\sigma}^*(\rbf )\psi_{\nu\sigma^\prime}(\rbf) = S_{\mu\nu}\delta_{\sigma\sigma^\prime}\,.
\end{equation}
Thus, the overlap matrix among the new basis functions is
\begin{eqnarray}
    \int d\rbf^3 \tilde\psi_{\mu_1\sigma_1}^*(\rbf)\tilde{\psi}_{\mu_2 \sigma_2}(\rbf) 
    = X^\dagger SX\, \delta_{\sigma_1\sigma_2}\,.
\end{eqnarray}
By setting the requirement $X^\dagger SX= I$ and considering $S$ being positive-definitive, a standard choice\cite{szabo2012modern} is $X = S^{-1/2}$.
This selection results in a new orthonormal basis set $\{\tilde{\psi}_{\mu\sigma}(\rbf)\}$. 

Representing the many-body wavefunction in the Fock space, spanned by the Slater determinants of $\{\tilde{\psi}_{\mu\sigma}(\rbf)\}$, we obtain the the second quantization of the many-body states\cite{helgaker2014molecular}:
\begin{eqnarray}
|\{i_k\alpha_k\sigma_k\}\rangle= c^\dagger_{i_N\alpha_N\sigma_N} \cdots c^\dagger_{i_2\alpha_2\sigma_2}c^\dagger_{i_1\alpha_1\sigma_1} |0\rangle\,,
\end{eqnarray} 
where subscripted ``$i$''s denote the site indices labeling the quantum dots; ``$\alpha$''s denote the orbital indices $(n,m)$; ``$\sigma$''s denote the spin index. 
To perform an exact diagonalization calculation, we further construct a second-quantized Hamiltonian, the generic form of which is 
\begin{equation} \label{eq:tighBindingHamiltonian}
\Ham = \Ham_{\rm non-int} + \Ham_{\rm int}\,.
\end{equation}
The non-interacting part $\Ham_{\rm non-int}$ corresponds to the hopping of an electron across orbitals and wells, while the interacting part $\Ham_{\rm int}$ contains all the possible interactions between multiple electrons. 
In the following subsections, we introduce the methodology and approximations to evaluate these two parts of Hamiltonians.


\subsection{Non-Interacting Part of Hamiltonian}
The non-interacting part of the Hamiltonian is quadratic in fermionic operators
\begin{equation} \label{eq:noninteractingHamiltonian}
\Ham_{\rm non-int} = \sum_{i\alpha\sigma}\sum_{j\beta\sigma} t_{ij\atop \alpha\beta} c_{i\alpha\sigma}^\dagger c_{j\beta\sigma}\,.
\end{equation}
For any Fock state, this quadratic term can be evaluated using single-electron states. To simplify the calculation, we can first evaluate the matrix elements using the original non-orthogonal basis $\{\psi_{\mu\sigma}(r)\}$, obtaining a matrix $h_{\mu\nu}$, which we can transform into the orthonormal basis. This results in
\begin{eqnarray}\label{eq:atomicOrbitalTTerms}
h_{\mu\nu}&=& \int d\rbf^3 \psi_{\mu\sigma}^*(\rbf)\left[ -\frac{\hbar^2}{2m_e^\star} \nabla^2 + \sum_i V(\rbf - \textbf{R}_i)\right]\psi_{\nu\sigma}(\rbf)\nonumber\\
&=& \varepsilon_\nu + \int d\rbf^3 \psi_{\mu\sigma}^*(\rbf)\sum_{j\neq i_\nu} V(\rbf - \textbf{R}_j)\psi_{\nu\sigma}(\rbf)\,,
\end{eqnarray}
for any spin $\sigma$.
The diagonal terms of $h_{\mu\nu}$ define the site energies associated with each (single-well) orbital $\nu$. Note, that this energy is not equal to the bare eigen-energy $\varepsilon_\nu$ in a single well, since the second term also has a finite diagonal contribution. This is an analog of the ``crystal field''. The off-diagonal terms in $h_{\mu\nu}$ define the hybridization between different orbitals.

The transformation into orthonormal basis is done by substituting Eq.~\eqref{eq:orthTransform} into \eqref{eq:atomicOrbitalTTerms}, resulting in
\begin{eqnarray}\label{eq:defTTerms}
t_{\mu\nu}&=&\langle i\alpha\sigma |\Ham_{\rm non-int} |j\beta\sigma\rangle\nonumber\\
 &=& \sum_{\mu^\prime\nu^\prime}\int d\rbf^3 X^*_{\mu^\prime\mu}\psi_{\mu^\prime\sigma}(\rbf)\Ham
 \psi_{\nu^\prime\sigma}(\rbf)X_{\nu^\prime\nu}\,.
\end{eqnarray}
Here $t_{\mu\nu}$ defines the site energy (diagonal) and hybridization (off-diagonal) of the orthonormal orbitals, which appears in Eq.~\eqref{eq:tighBindingHamiltonian}. Due to the impact of the ``crystal field'' and hybridization, the energy distribution of a multi-well system can be dramatically different from the single-well solution.

\begin{figure}[!b]
\begin{center}
\includegraphics[width=\columnwidth]{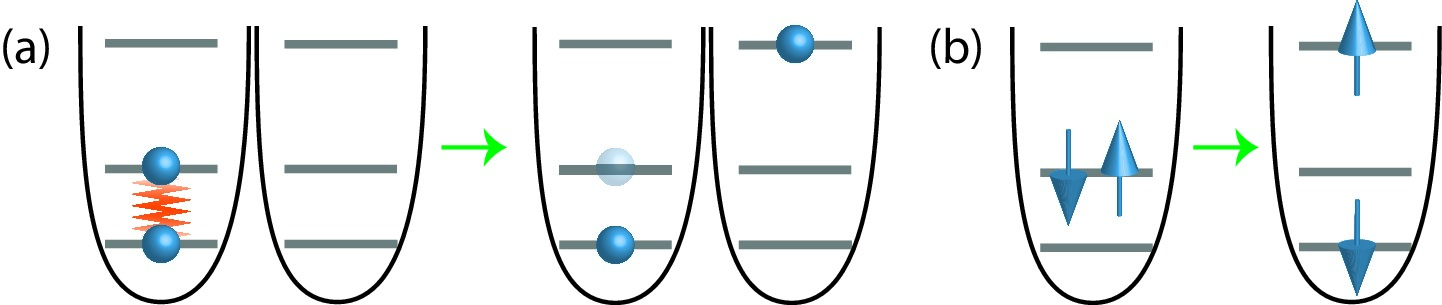}
\caption{\label{fig:multipletexcludedTerms} The interaction terms that are ignored in the tight-binding Hamiltonian: (a) the density-dependent hopping and (b) the scattering terms involving more than two orbitals with different site energies.}
\end{center}
\end{figure}

\subsection{Interacting Part of Hamiltonian}\label{sec:interactiongHam}
As a typical choice in condensed matter, we restrict the interaction part $\Ham_{\rm int}$ to four-fermion terms\cite{szabo2012modern}. As specified in Eq.~\eqref{eq:genericFourFermionTerms}, the generic second-quantized four-fermion term contains an enumeration of four coordinate indices ($i_1,i_2,j_1,j_2$), four orbital indices ($\alpha_1, \alpha_2, \beta_1, \beta_2$) and four spin indices ($\sigma_1, \sigma_2, \sigma^\prime_1,\sigma^\prime_2$). This results in $16N_{\rm well}^4 N_{\rm orbital}^4$ interaction terms, whose general expression is shown in Eq.~\eqref{eq:generalVterms}. The bottleneck of the computation is the evaluation of the interaction parameters by numerical integration. Therefore, the setup of a model involving all combinatorial possibilities is currently beyond our capability. Hence, we introduce several common approximations to reduce the number of independent variables. Firstly, without relativistic effects, the Coulomb interaction is independent of spin; therefore, $\{\sigma^\prime_1,\sigma^\prime_2\} = \{\sigma_1,\sigma_2\}$. Secondly, due to the two-body nature of the interaction, one-center and two-center integrals dominate, whereas terms with more centers decay exponentially for well-separated wells. Dropping these multi-center terms implies the assumption that the geometric coordinates \{$i_1$, $i_2$, $j_1$, $j_2$\} can take at most two values. Thirdly, to further reduce the complexity, we restrict the interaction terms to ``perfectly'' resonant processes\cite{resonance}, which strictly speaking is only fully justified when level splitting is much larger than the interaction energy scales. For example, we neglect two generic classes of interactions: the density-dependent hopping and the scattering terms involving more than two orbitals [see Fig.~\ref{fig:multipletexcludedTerms}]. These terms are important in some cold-atom systems where individual energy scales are controllable, but become non-resonant in our model due to the strong interaction and unequal spacing between energy levels\cite{nonresonance}. The omission of these two non-resonant processes is equivalent to setting $\{i_1,i_2\}=\{j_1,j_2\}$ and restricting each four-fermion interaction term to at most two orbital indices. The above approximations significantly reduce the complexity of the model and have been a common strategy in solid state\cite{dagotto2013nanoscale}.

\begin{figure}[!t]
\begin{center}
\includegraphics[width=\columnwidth]{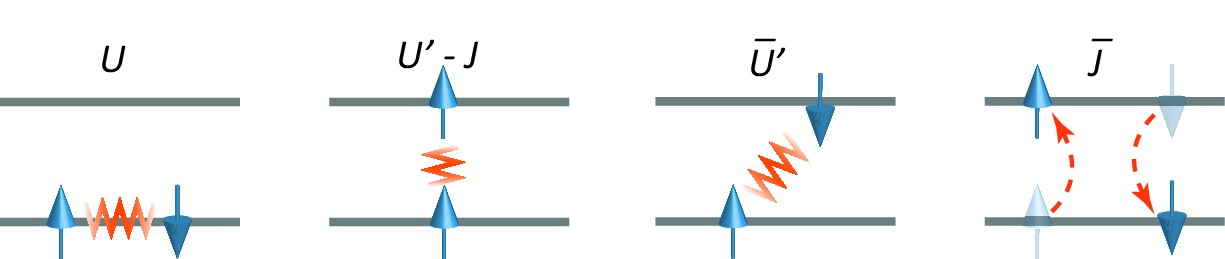}
\caption{\label{fig:multipletOnsite} On-site interactions within one quantum dot: Hubbard $U$, inter-orbital Hubbard $U^\prime$ (and its spin-anti-parallel form $\bar{U}^\prime$), and Hund's exchange $J$ (and its spin-anti-parallel form $\bar{J}$).}
\end{center}
\end{figure}

After these simplifications, the interacting part of the Hamiltonian can be decomposed as the on-site and (two-center) long-range parts
\begin{equation} \label{eq:decomposingIntHamiltonian}
\Ham_{\rm int} = \sum_i \Ham_i^{\rm(OS)} + \sum_{ij} \Ham_{ij}^{\rm(LR)}.
\end{equation}
The standard derivations of each term in the interacting Hamiltonian is present in Appendix~\ref{app:interactiongHam}. 
The on-site interaction Hamiltonian can be written as
\begin{eqnarray}\label{eq:multipletHam}
\Ham^{\rm (OS)}_i\!&=&\!\frac12\sum_{\alpha\sigma} U_{\alpha} n_{\alpha\bar{\sigma}}n_{\alpha\sigma} + \frac12\!\sum_{\alpha_1\neq\alpha_2}\!\sum_{\sigma_1,\sigma_2} U^\prime_{\alpha_1 \alpha_2} n_{\alpha_2\sigma_2}n_{\alpha_1\sigma_1} \nonumber\\
&& +  \frac12\sum_{\alpha_1\neq\alpha_2}\sum_{\sigma_1,\sigma_2} J_{\alpha_1 \alpha_2}
 c^\dagger_{\alpha_2\sigma_1}c^\dagger_{\alpha_1\sigma_2}  c_{\alpha_2\sigma_2}c_{\alpha_1\sigma_1} \,.
\end{eqnarray}
This is the known as the \textit{multiplet model}, widely used to describe the valence electrons in the transition metal systems\cite{dagotto2013nanoscale}.
The corresponding scattering processes are sketched in Fig.~\ref{fig:multipletOnsite}. For convenience, the site index is removed on the right-hand side, while in an inhomogeneous system [such as the modulations in Sec.~\ref{sec:calculations}], one should consider it specifically for each individual site.
Due to a symmetry consideration [see discussions in Appendix~\ref{app:integral}], it is usually convenient to calculate the interaction parameters using the original single-well basis obtained from Eq.~\eqref{eq:singleWellBasis}, through
\begin{equation}\label{eq:expXi}
    \Xi_{\mu\!_1\nu\!_1\atop\mu\!_2\nu\!_2}\!=\! \iint\! d\rbf_1^dd\rbf_2^d\, W(|\rbf_1\! -\!\rbf_2|)\psi_{\mu_1}\!(\!\rbf_1\!)^*\psi_{\mu_2}\!(\!\rbf_2\!)^*\psi_{\nu_1}\!(\!\rbf_1\!)\psi_{\nu_2}\!(\!\rbf_2\!)\,.
\end{equation}
Here $W(|\rbf_1\! -\!\rbf_2|)={e^2}/{4\pi\epsilon|\rbf_1 -\rbf_2|}$ is the two-electron Coulomb repulsion.
Note, that here we have taken the compact notation $\mu = (j,\beta)$ introduced above and have omitted the spin indices as they do not affect the spatial integral. Then using Eq.~\eqref{eq:orthTransform}, we have 
\begin{equation}\label{eq:onsiteParams}\begin{aligned}
    U^\prime_{\alpha_1 \alpha_2} &=\sum_{\mu_1,\mu_2}\sum_{\nu_1,\nu_2}X_{\mu_1a_1}^*X_{\nu_1 a_1}X_{\mu_2a_2}^*X_{\nu_2a_2}\Xi_{\mu_1\nu_1\atop\mu_2\nu_2}\\
    J_{\alpha_1 \alpha_2} &= \sum_{\mu_1,\mu_2}\sum_{\nu_1,\nu_2}X_{\mu_1a_2}^*X_{\nu_1 a_1}X_{\mu_2a_1}^*X_{\nu_2a_2}\Xi_{\mu_1\nu_1\atop\mu_2\nu_2}.
\end{aligned} \end{equation}
These parameters define the on-site interactions among the orthonormal orbitals.

\begin{figure}[!t]
\begin{center}
\includegraphics[width=0.95\columnwidth]{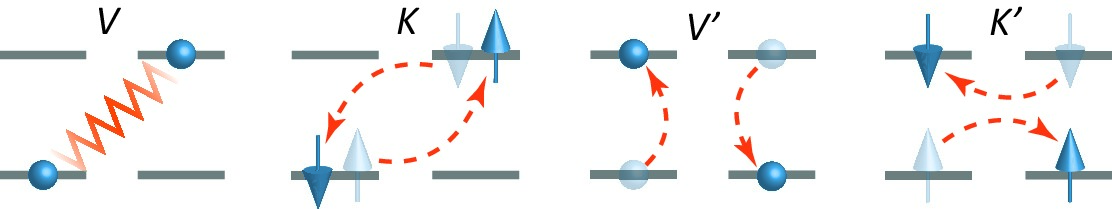}
\caption{\label{fig:multipletOffsite} Long-range interactions between two quantum dots: direct Coulomb interaction $V$, long-range Hund's exchange $K$, correlated on-site exchange $V^\prime$ and correlated off-site exchange $K^\prime$.
}
\end{center}
\end{figure}

Similarly, the long-range interactions are written as 
\begin{eqnarray}\label{eq:longrangeHam}
\Ham^{\rm (LR)}_{ij}&=&\frac12\sum_{\alpha\sigma}\sum_{\beta\sigma^\prime} V_{\alpha\beta} n_{i\alpha\sigma}n_{j\beta\sigma^\prime}\nonumber\\
&& + \frac12\sum_{\alpha\beta}\sum_{\sigma\sigma^\prime} K_{\alpha\beta}
 c^\dagger_{j\beta\sigma}c^\dagger_{i\alpha\sigma^\prime}  c_{j\beta\sigma^\prime}c_{i\alpha\sigma}\nonumber\\
 &&+ \frac12\sum_{\alpha\neq\beta}\sum_{\sigma\sigma^\prime} V^\prime_{\alpha\beta}
 c^\dagger_{i\beta\sigma}c^\dagger_{j\alpha\sigma^\prime}  c_{j\beta\sigma^\prime}c_{i\alpha\sigma}\nonumber\\
 && + \frac12\sum_{\alpha\neq\beta}\sum_{\sigma\sigma^\prime} K^\prime_{\alpha\beta}
 c^\dagger_{j\alpha\sigma}c^\dagger_{i\beta\sigma^\prime}  c_{j\beta\sigma^\prime}c_{i\alpha\sigma}\nonumber\\
 && + \frac12\sum_{\alpha\neq\beta}\sum_{\sigma\neq\sigma^\prime} K^{\prime\prime}_{\alpha\beta}
 c^\dagger_{i\beta\sigma}c^\dagger_{j\beta\sigma^\prime}  c_{j\alpha\sigma^\prime}c_{i\alpha\sigma}\,.
\end{eqnarray}
As sketched in Fig.~\ref{fig:multipletOffsite}, $V_{\alpha\beta}$ represents a direct Coulomb interaction and $K_{\alpha\beta}$ is the corresponding exchange interaction; similarly, $V^\prime_{\alpha\beta}$ is the correlation between two on-site exchange interactions, while $K^\prime_{\alpha\beta}$ is the correlation between off-site exchange. The $K^{\prime\prime}_{\alpha\beta}$ term is an analog of the pair-hopping term and is also ignored here.
The expressions for the relevant long-range terms are
\begin{eqnarray}\label{eq:longrangeParams}
    V_{\alpha\beta}^{ij} &=& \sum_{\mu_1,\mu_2}\sum_{\nu_1,\nu_2}X_{\mu_1(i\alpha)}^*X_{\nu_1(i\alpha)}X_{\mu_2(j\beta)}^*X_{\nu_2(j\beta)}\Xi_{\mu_1\nu_1\atop\mu_2\nu_2}\nonumber\\
    K_{\alpha\beta}^{ij} &=& \sum_{\mu_1,\mu_2}\sum_{\nu_1,\nu_2}X_{\mu_1(j\beta)}^*X_{\nu_1 (i\alpha)}X_{\mu_2(i\alpha)}^*X_{\nu_2(j\beta)}\Xi_{\mu_1\nu_1\atop\mu_2\nu_2}\nonumber\\
    V_{\alpha\beta}^{ij\prime} &=& \sum_{\mu_1,\mu_2}\sum_{\nu_1,\nu_2}X_{\mu_1(i\beta)}^*X_{\nu_1(i\alpha)}X_{\mu_2(j\alpha)}^*X_{\nu_2(j\beta)}\Xi_{\mu_1\nu_1\atop\mu_2\nu_2}\nonumber\\
    K_{\alpha\beta}^{ij\prime} &=& \sum_{\mu_1,\mu_2}\sum_{\nu_1,\nu_2}X_{\mu_1(j\alpha)}^*X_{\nu_1 (i\alpha)}X_{\mu_2(i\beta)}^*X_{\nu_2(j\beta)}\Xi_{\mu_1\nu_1\atop\mu_2\nu_2}\,.
\end{eqnarray}
Note, the long-range interaction has contributions from both direct long-range integrals (for two-center $\mu_i$ and $\nu_i$ indices), and indirect hybridized on-site integrals (for one-center $\mu_i$ and $\nu_i$ indices). With well-separated quantum dots, the long-range interactions are typically much smaller than the on-site interactions. That being said, $V\ll U$, $K\ll J$, and $V^\prime$ and $K^\prime$ are even smaller compared to $V$ and $K$. Due to the orbital match of on-site wavefunctions, the $V$ terms are expected to dominate in the long-range interactions. However, for the study of Nagaoka ferromagnetism in the plaquette (see Sec. ~\ref{sec:calculations}), it is necessary to consider all of these long-range parameters, since the effects we want to observe can depend significantly on the superfine structures.

\begin{figure}[!t]
\begin{center}
\includegraphics[width=\columnwidth]{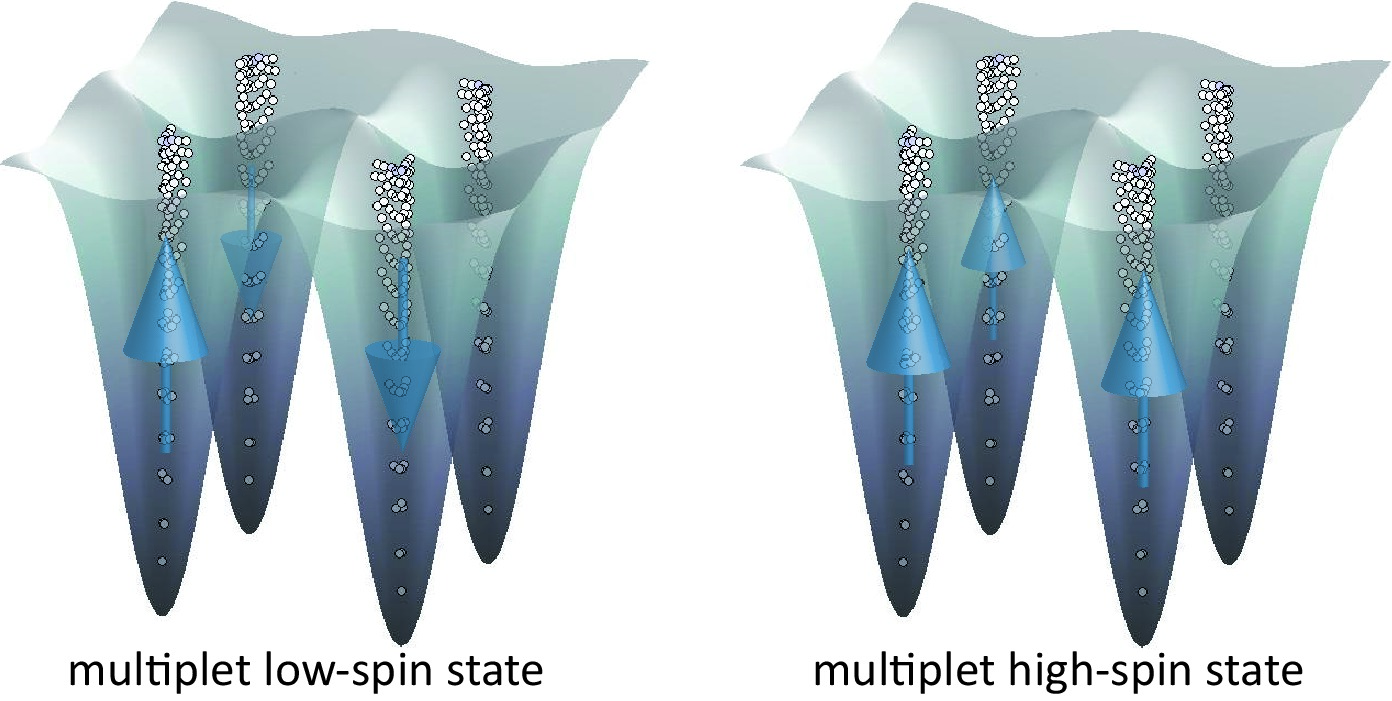}
\caption{\label{fig:multiplet} Cartoon for multi-orbital Nagaoka transition in a four-dot system. For moderate effective interaction, the multiplets in each quantum well form an overall low-spin state, with total spin $S=1/2$ (left). In contrast, a large interaction relative to the tunneling gives a Nagaoka FM state (right). The shaded surfaces denote the potential wells, while the white dots denote the single-well energy levels, which are slightly shifted according to different angular quantum numbers. The spin configuration is a conceptual sketch instead of a realistic solution.
}
\end{center}
\end{figure}
\section{Simulation of Four-Well Quantum Dot System: Probing Nagaoka Magnetism}\label{sec:calculations}
The explicit expressions for the tight-binding parameters described above allow one to fully diagonalize many-body electronic systems with multiple quantum dots. We will use this methodology to investigate the physics described by Nagaoka\cite{nagaoka1966ferromagnetism}, applied to a multi-orbital, 2$\times$2 system. Specifically, we study a system with three electrons in a four-site plaquette, which realizes the condition of a single hole in a Mott insulator where for a single orbital per site Nagaoka proved that the ground state must be ferromagnetic in the limit of large interaction strength. As sketched in Fig.~\ref{fig:multiplet}, with the total electron occupation less than the number of quantum dots, the multiplets on each quantum dot interact with each other and are expected to yield an effective collective spin configuration. If the multi-orbital system has similar behaviors to those of a single-band system formed by those multiplets, we expect it to display a high-spin-low-spin transition at various model parameters: with large enough interaction relative to the tunneling, we expect the Nagaoka mechanism to yield a ferromagnetic (FM) high-spin ground state; however, with moderate interactions, the system becomes a doped Mott insulator with a low-spin ground-state configuration, which corresponds to an anti-ferromagnetic state in the thermodynamic limit\cite{von2010probing}.

A recent experiment has studied Nagaoka magnetism using a quantum dot array in a $2\times2$ plaquette configuration\cite{dehollain2019nagaoka}. For a great part of the analysis in that work, a single-band extended Hubbard model with fitted parameters was used to model the system, obtaining results that seem to describe most of the experimental observations accurately. However, the fact that the experimentally observed level spacing between the two lowest orbitals is smaller than the electronic interaction raises the question of whether the system is adequately described by the single-band model. In this section, we use the \emph{ab initio} exact diagonalization approach described above to extract the precise many-body model of the $2\times2$ quantum dot plaquette and quantitatively reproduce the Nagaoka conditions that were explored with the experimental system.

\subsection{Evaluation of Model Parameters}
To compare with a realistic system, we first discuss the typical values of parameters. The gate-electrode structure of the experimental device was lithographically designed to define quantum dot wells on the scale of 100\,nm\cite{mukhopadhyay20182,dehollain2019nagaoka}. Therefore, we set our spatial units of the lattice constant $a_0=100$\,nm and Gaussian potential width $\Sigma = 100$\,nm. Considering the effective mass of electrons in a GaAs/AlGaAs  2DEG is $m_e^\star \approx 0.067\,m_e$, the natural energy unit corresponds to ${\hbar^2}/{a_0^2m_e^\star}\approx 0.114$\,meV. Applying this scale to the eigen-spectrum solved in Fig.~\ref{fig:1} ({i.e.},~$V_0 = 100\, {\hbar^2}/{a_0^2m_e^\star} = 11.4$\,meV), we obtain the ground state to first excited state level spacing $\Delta E = \varepsilon_1-\varepsilon_0 \approx 0.75$\,meV, comparable to the experimental observation of $\sim 1$\,meV.

\begin{figure}[!b]
\begin{center}
\includegraphics[width=\columnwidth]{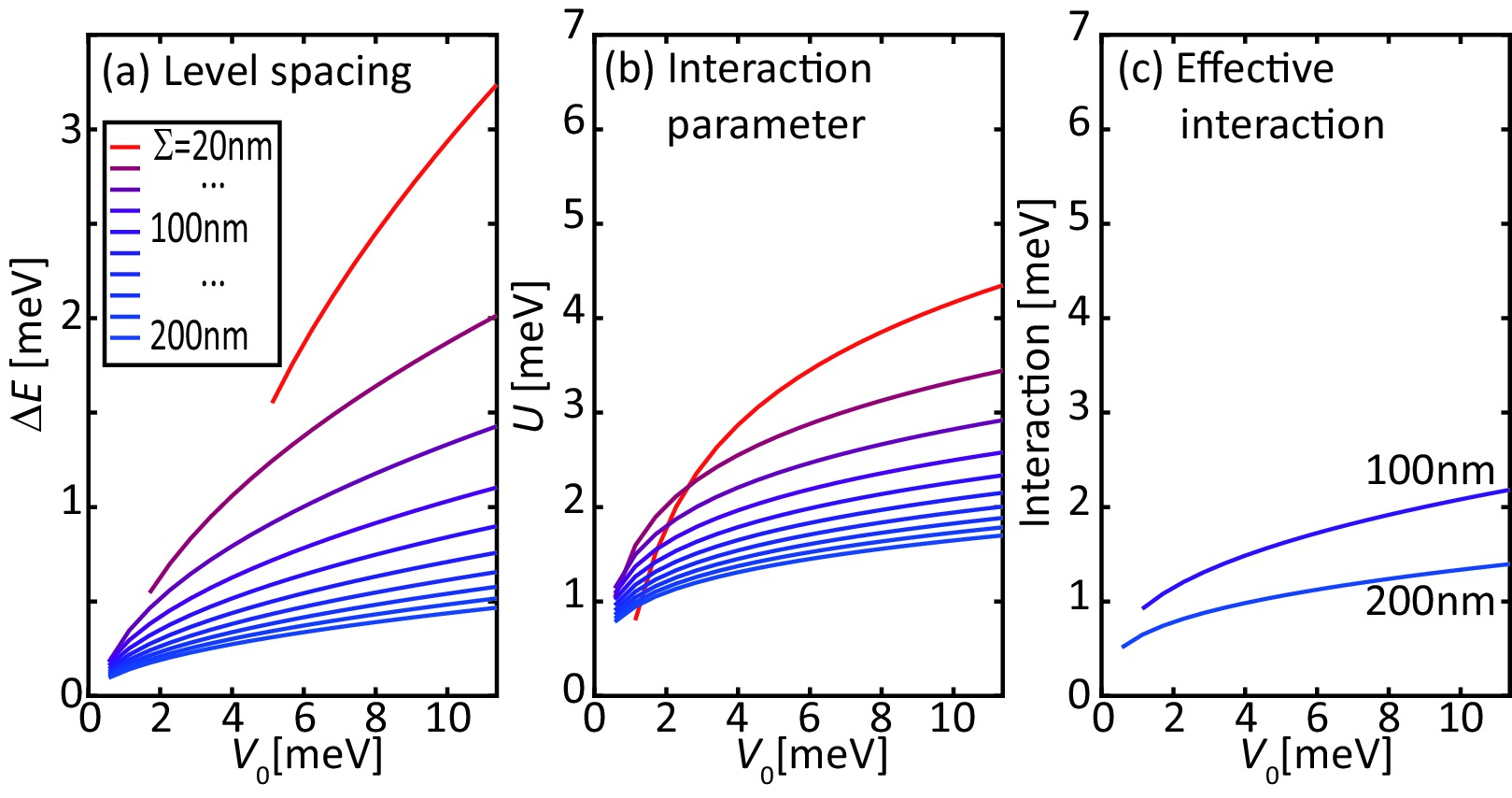}
\caption{\label{fig:interaction} (a) Ground-first-excited state level spacing, (b) ground-state interaction $U$ and (c) effective interaction as a function of the depth $V_0$ and the width $\Sigma$ of quantum well. The calculation is obtained on a single quantum well without hybridization.
} 
\end{center}
\end{figure}

The evaluation of the electron-electron interaction requires a specific value of the dielectric constant $\epsilon$, which is ideally 12.9 in GaAs. It is known that the presence of metallic gate electrodes in the vicinity of the 2DEG has the effect of increasing $\epsilon$. However, the precise evaluation of $\epsilon$ is challenging. Instead, we rely on the value of the addition energy, which has been accurately estimated by experiments to be 2.9 meV, and select an $\epsilon$ that results in reasonable interaction values. Taking $\epsilon = 20$ into the solution of $V_0 = 11.4$\,meV mentioned above gives the ground-state $U_0\approx2.34$\,meV and the ground-excited-state $U^\prime_{01} \approx 1.92$\,meV. Note, these are the intrinsic model parameters in the many-body Hamiltonian. A typical experimental estimation of this Hubbard interaction is obtained by measuring the addition energy. Due to the orbital mixture when $\Delta E < U$ and the fact that excited-state wavefunctions are spatially wider, the experimentally measured ``effective interaction'' strength is slightly smaller than the model parameters $U$ and $U^\prime$. Figure~\ref{fig:interaction} gives an example of level spacing $\Delta E$, ground-state Hubbard $U$ and effective interaction calculated in a single-well system with different shape parameters.

The long-range interactions are much smaller than the on-site ones. Specifically for $d=210$\,nm, the long-range interaction $V$ ranges from 0.22\,meV to 0.4\,meV depending on the orbitals; $K$ and $V^\prime$ are on the order of or below 1\,$\mu$eV; the off-site exchange correlation $K^\prime$s are even lower, on the order of 0.1 or 0.01\,$\mu$eV. These terms form higher-order corrections to the multiplet model of Eq.~\eqref{eq:multipletHam}. As shown in Table \ref{tab:modelComp}, only the long-range Coulomb interaction $V$ obviously affects the ground-state energy, by order of 1 meV, while others contribute to $\sim0.01$\,meV. However, as stated before and now made clear in Table \ref{tab:modelComp}, the strong interaction condition results in a high-spin to low-spin state energy gap--which we refer to as the Nagaoka gap, that is on the scale of $\mu$eV. The precise value of the Nagaoka gap depends on the details of the microscopic parameters such as the confining potential for electrons and the many-body interactions. Therefore, every long-range term provides a non-negligible contribution to the Nagaoka gap. Noticeably, the $K$ terms have larger contributions to the Nagaoka gap than $V$, although it is inconsistent to include only one of them because it is the combination of both that obeys the exchange relation in Eq.~\eqref{eq:generalVterms}. A closer inspection of the dependence of the Nagaoka gap size on various models -- in particular the contrast between $t$-$U$-$J$ and $t$-$U$-$J$-$V$-$K$-$V^\prime$-$K^\prime$ -- indicates that the long-range Hund's exchange only contributes $\sim 23\%$ of the ferromagnetic effect, with the Nagaoka mechanism dominating. Distinguishing these two contributions is only possible in a multi-band model. This quantitative assignment gives further confirmation that the experimental result in Ref.~\onlinecite{dehollain2019nagaoka} is indeed caused by a Nagaoka-like mechanism.

\begin{table}
    \begin{tabular}{c|c|c}
    \hline \hline
 Model& Ground-State Energy  & Nagaoka Gap  \\
 \hline 
 $t$-$U$-$J$  & -43.579950\,meV & 2.213\,$\mu$eV \\
 $t$-$U$-$J$-$V$ & -42.576572\,meV & 2.318\,$\mu$eV\\
 $t$-$U$-$J$-$V$-$K$ & -42.558866\,meV & 2.775\,$\mu$eV \\
 $t$-$U$-$J$-$V$-$K$-$V^\prime$-$K^\prime$ & -42.558912\,meV & 2.868\,$\mu$eV\\
 \hline  \hline
\end{tabular}
\caption{\label{tab:modelComp} Effect of system parameters (definition of these parameters can be found in the Sec.~\ref{sec:interactiongHam}) on ground-state energies and the Nagaoka gap obtained by various models for $d=210$\,nm. The calculations are performed on a four-dot system with three electrons, and the ground states of all models listed here are high-spin states.}
\end{table}

\begin{figure}[!t]
\begin{center}
\includegraphics[width=\columnwidth]{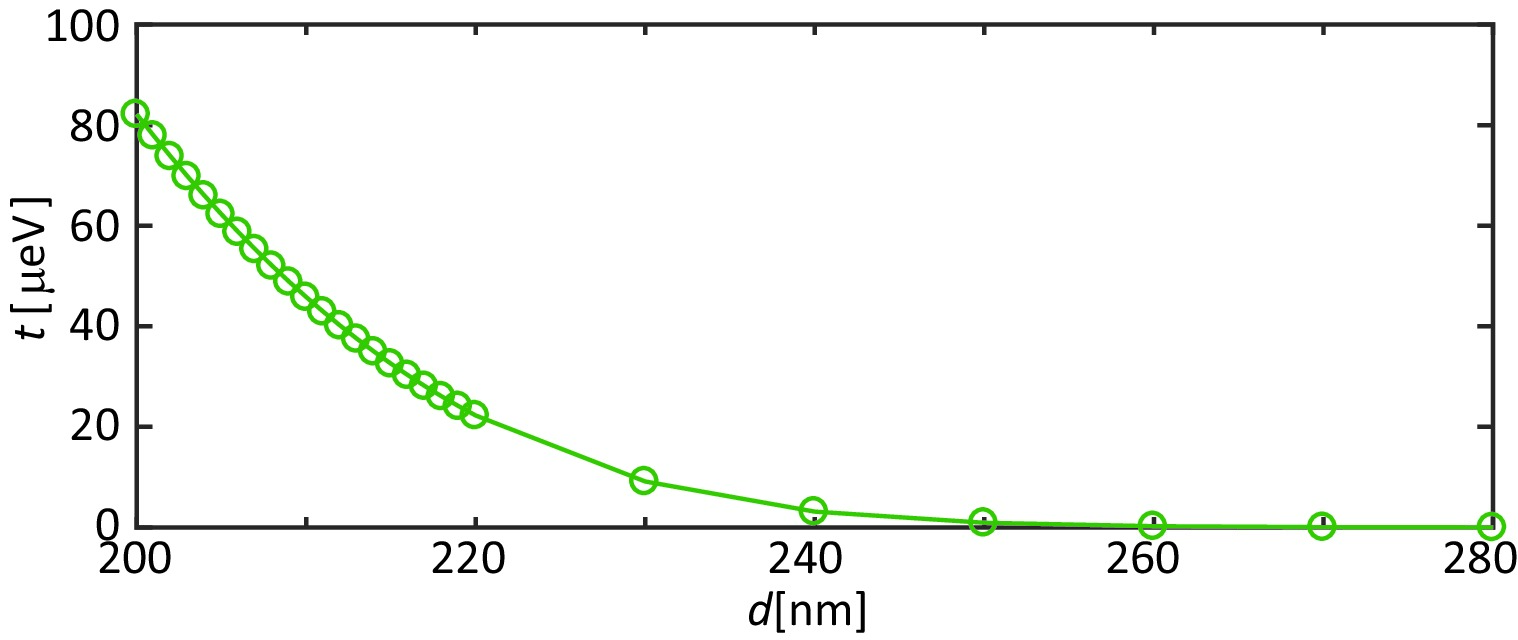}
\caption{\label{fig:bandwidth} The effective hopping $t$ estimated by a quarter of the single-particle bandwidth calculated for various distances in a $2\times 2$ plaquette.
}
\end{center}
\end{figure}
The hybridizations, or tunneling terms, vary among different orbitals and are exponentially dependent on the distance between quantum dot potential wells. Since the single-well ground-state wavefunctions are most localized, the hybridizations between neighboring-quantum-dot (single-well) ground states are extremely small ($\sim0.06\,\mu$eV for $d=210$\,nm). However, these local orbitals and tunnelings among them $h_{\mu\nu}$ are non-physical: they are nothing but mathematical tools to solve the many-body problem\cite{coherence}. In reality, the ``crystal field'' and wavefunction orthogonalization cause heavy hybridization between the (single-well) ground state and excited states -- the maximum of which can be close to $\Delta E$. These high-level excited states can contribute a $\sim 0.5$\,meV hopping amplitude between neighboring quantum wells. Therefore, the experimentally measurable \emph{effective tunneling} across low-energy states is the result of a superposition of all different conceptual paths.

Following this philosophy, the effective hopping $t$ can be simply extracted from the single-particle bandwidth in the entire multi-well system. If we only consider nearest-neighbor tunneling, the low-energy band structure of a $2\times 2$ plaquette takes the form $E(\theta) = -2t\cos \theta$ for $\theta = \{0, \pi/2, \pi, 3\pi/2\}$. Therefore, the width of the lowest band (the lowest four states) in a single-electron system gives an estimation of $4t$. Figure~\ref{fig:bandwidth} shows the extracted values of $t$ for different neighboring-dot distances. In the experimental device, the inter-dot tunneling can be tuned to the range of 0-40\,$\mu$eV\cite{mukhopadhyay20182, dehollain2019nagaoka}, which in the \emph{ab initio} model corresponds to a range of distances $d=210-240$\,nm. This is fairly consistent with the lithographically designed inter-dot distance of 150\,nm, which is also an approximation, since the actual inter-dot distance in the experiments is not measurable.

We emphasize that the above model parameters (including the first excited-state level spacing $\Delta E$, the ground-state and ground-excited-state Coulomb interaction $U_0$ and $U_{01}^\prime$, long-range Coulomb interaction $V$, the effective tunneling $t$, and the Nagaoka gap $\Delta$) evaluated from our \emph{ab initio} calculation using only very limited experimental input  match quantitatively with the experiment in Ref.~\onlinecite{dehollain2019nagaoka}. Therefore, we believe the \emph{ab initio} calculation serves the purpose of predicting model parameters in a quantitative level based only on given potential landscapes.

To simulate the correlated Nagaoka physics in multiple quantum dots, we perform the calculation in a microcanonical ensemble, with three electrons in a four-well system, and focus on the ground-state properties. The evaluation of single-well eigenstates and the integration are performed on a grid with a spacing of 1\,nm. To simplify the calculation, we keep 15 orbitals in each quantum dot, which span a $\sim 5$\,meV energy range. As this range is much larger than both $U$ and $t$, we believe that the level mixture above this truncation can be ignored\cite{convergence}. We perform exact diagonalization to solve this 60-orbital spinful system, using the parallel Arnoldi approach\cite{lehoucq1998arpack, jia2017paradeisos}.

\begin{figure}[!t]
\begin{center}
\includegraphics[width=\columnwidth]{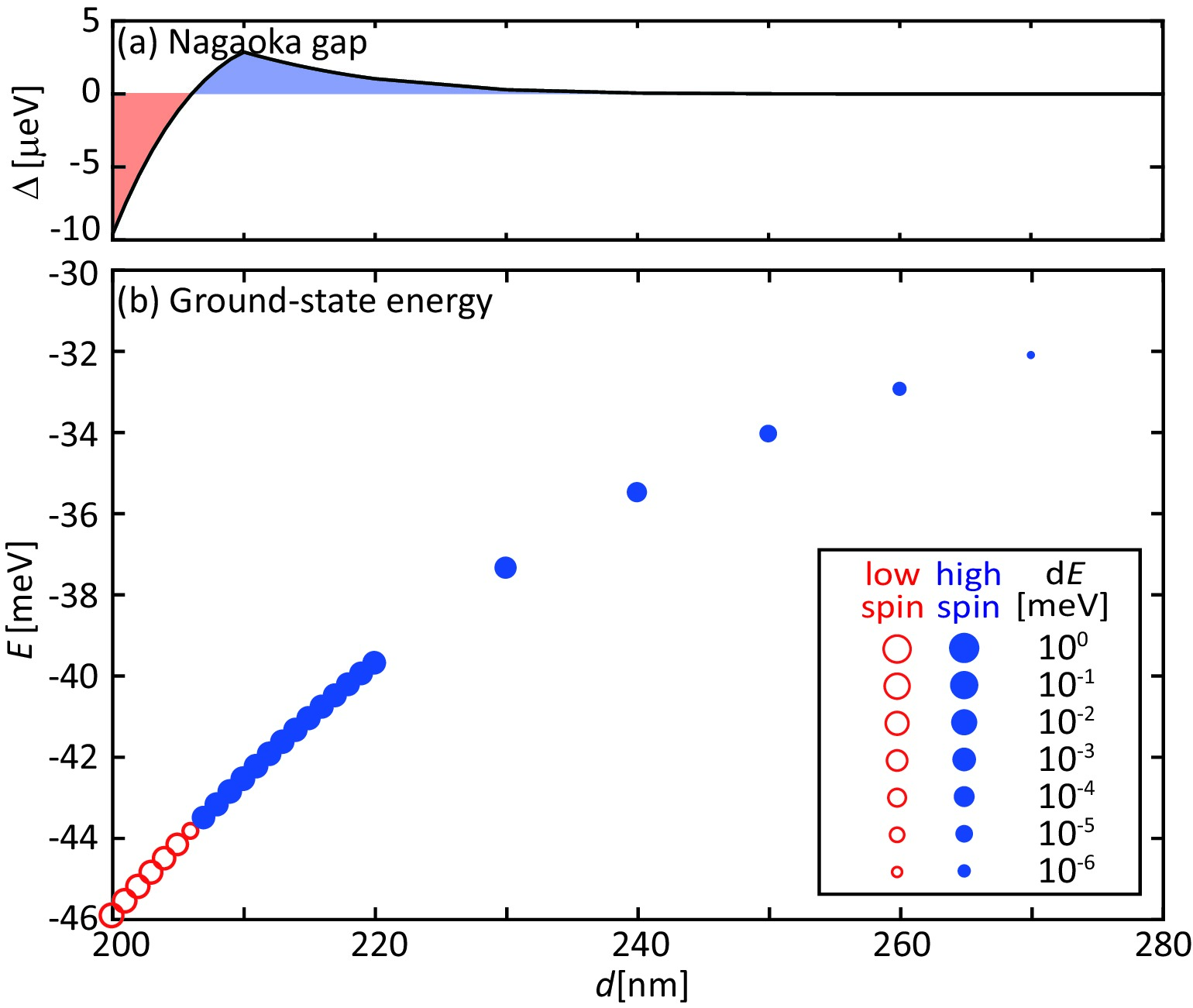}
\caption{\label{fig:distDep} (a) The Nagaoka gap and (b) the ground-state energy of three electrons in four quantum dots, as a function of the distance $d$. The red open circles denote the low-spin ground states, while the blue dots denote the high-spin ground states. The size of the data points reflects the energy difference between the lowest low-spin and high-spin states in a logarithmic scale.
}
\end{center}
\end{figure}
\subsection{Distance Dependence}
Having selected the quantum dot potential well parameters, we first study the ground state properties as a function of the distance between neighboring dots in the plaquette. As shown in Fig.~\ref{fig:distDep}(b), the energy increases monotonically when the quantum dot separation is increased from 200\,nm to 280\,nm. This is a consequence of the crystal field renormalization of the site energies. As the dot separation becomes large enough to make the long-range interactions negligible, the electrons can no longer lower their energy by delocalizing, and the ground-state energy saturates towards $\sim 30$\,meV. This energy corresponds to each of the three electrons occupying the ground state of a quantum well independently.

Interestingly, the ground-state configuration switches from a high- to a low-spin state at $d\gtrsim 206$\,nm. This is a feature of the Nagaoka effect applied to finite-size lattices, which have access to regimes outside of the thermodynamic limit ($U/t \rightarrow \infty$) where Nagaoka made the original prediction. Increasing the distance between dots effectively suppresses $t$ and long-range interactions, but changes little of the on-site interactions. At small enough effective tunneling with large enough distance, the $U\gg t$ condition is reached at some point. Such a Nagaoka effect was originally predicted for a single hole in a half-filled Hubbard model in the thermodynamic limit, where the transition occurs at an infinite $U/t$ ratio. However, this critical ratio becomes finite for a finite cluster, since the underlying physics reflected by the Nagaoka transition is a $t$ versus $N_{\rm well}\times J$ competition. This phenomenon was previously shown (and proven) in a single-band Hubbard\cite{nagaoka1966ferromagnetism, mattis2003eigenvalues} and extended Hubbard models\cite{kollar1996ferromagnetism}. Here we show its validity in a multi-orbital system.

\begin{figure}[!b]
\begin{center}
\includegraphics[width=\columnwidth]{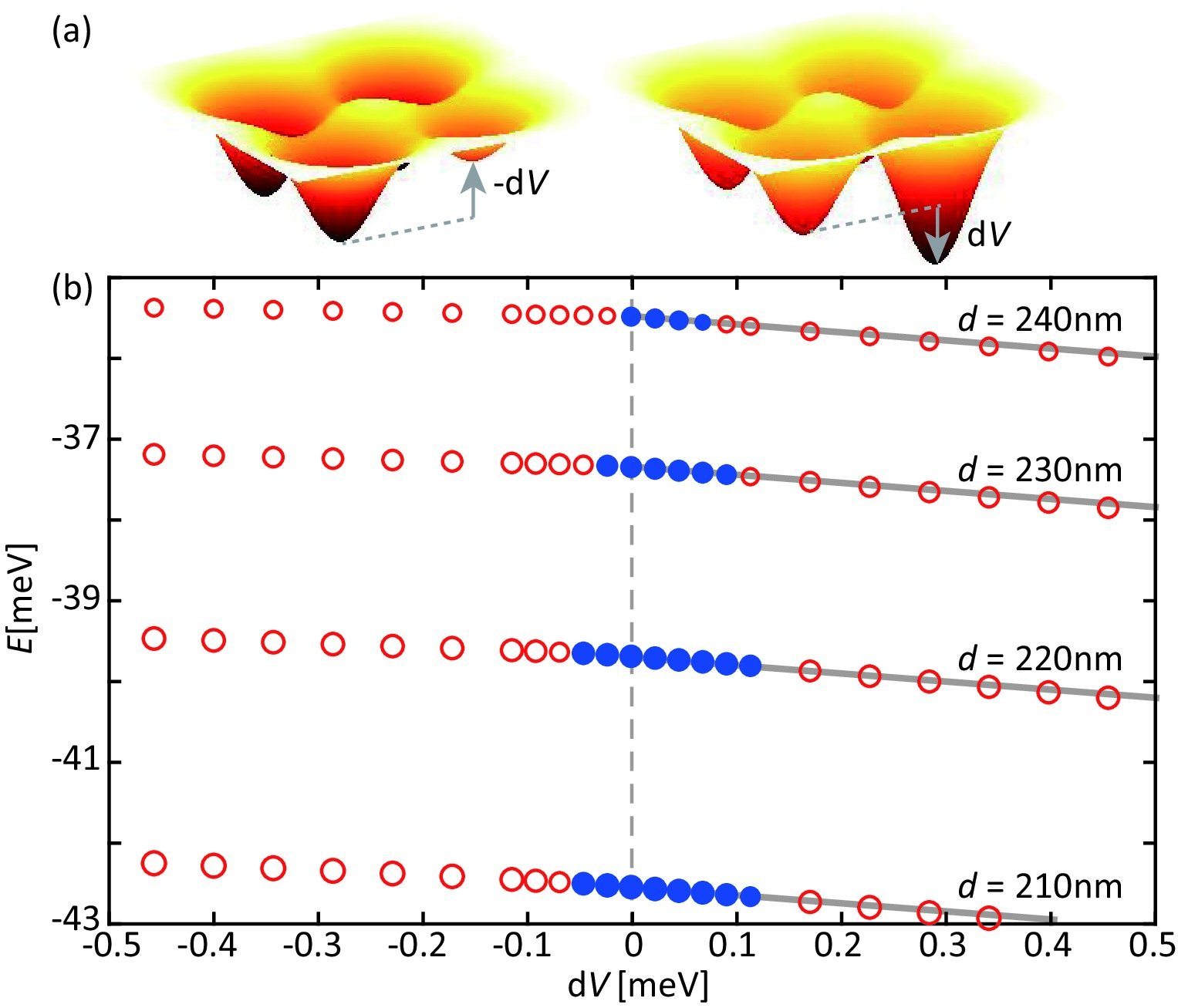}
\caption{\label{fig:detuningGroundState} (a) Schematic illustrating the potential detuning applied on one of the four quantum dot potential wells. (b) The ground-state energy for the entire system as a function of the potential detuning $dV$, calculated for various distances $d$. The gray lines denote the energy drop with slope 1.
}
\end{center}
\end{figure}

As shown in Fig.~\ref{fig:distDep}(a), the Nagaoka gap switches to positive at $d> 206$\,nm and reaches a maximum at $d\sim 210$\,nm. With larger distances, the Nagaoka gap starts to decrease as the correlations among electrons in different wells diminish. We select $d= 210$\,nm as the default geometric setup for the following calculations. In this case, the absolute value of the Nagaoka gap is 2.87\,$\mu$eV, consistent with an estimation in Ref.~\onlinecite{dehollain2019nagaoka} through a comparison between experimentally measured parameters with a fitted single-band model.

\subsection{Potential Detuning}

In addition to investigating the Nagaoka transition as a function of separation between the dots, we demonstrate that the low-spin-high-spin transition can also be driven by varying the potential of a single well, which reflects the robustness of the magnetism against disorder. As shown in Fig.~\ref{fig:detuningGroundState}(a), we vary the depth $V_0$ of one of the wells by a positive or negative $dV$, which results in unbalanced site-energies. More broadly, the change of all eigenstates associated with this particular well affects the hybridization and interaction parameters. These changes are all captured in the \emph{ab initio} calculation.

The results from this study, shown in Fig.~\ref{fig:detuningGroundState}(b), give some expected, but also some unexpected outcomes. A first observation is that the total energy of the system is lowered as the selected well is made deeper, and the Nagaoka condition breaks when the well becomes sufficiently shallow or deep. Surprisingly though, the slope of such energy decrease varies when $dV$ switches from positive to negative. Additionally, there is an asymmetry in the robustness of the Nagaoka state, between positive and negative detuning, which was also observed in the experiment\cite{dehollain2019nagaoka}. Taking the $d=210$\,nm system as an example, at $dV=0$, the ground state is the Nagaoka high-spin state discussed above; when the potential detuning is $dV\!=\!0.11$\,meV or $dV\!=\!-0.07$\,meV, the system undergoes a transition to the low-spin ground state. The asymmetric behaviors indicate that the transitions at positive and negative $dV$s have a different nature. 

\begin{figure}[!t]
\begin{center}
\includegraphics[width=\columnwidth]{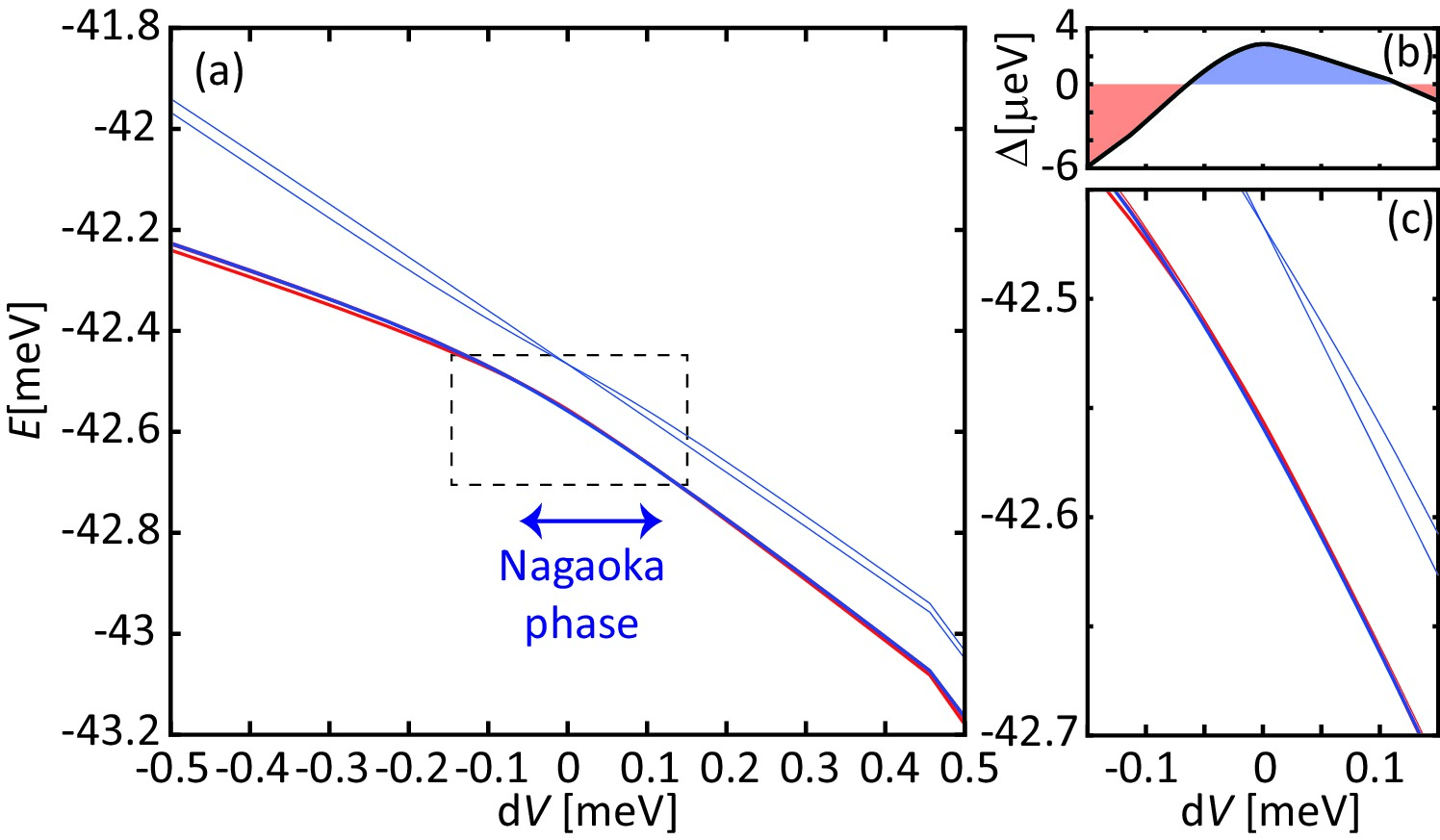}
\caption{\label{fig:detuningExcitedState} (a) The first three excited-state energies in high-spin (blue) and low-spin (red) sectors. The arrow denote the region of Nagaoka phase. (b) The Nagaoka gap and (c) an enlarged energy devolution for the dashed boxed region in (a).
}
\end{center}
\end{figure}

For $dV>0$, the detuned well is deeper, lowering the energy barrier for a doubly-occupied state (sometimes called doublon) and accordingly increasing the spin-exchange energy $J$ through the super-exchange process\cite{petta2005coherent,hanson2007spins}. Thus, the ground state becomes a low-spin state for large enough $dV$. We note that the range of $dV$ that we are sweeping is smaller than the Hubbard interactions (on the order of meVs); therefore, the transition is not caused by a direct doublon formation in the detuned site. In addition, the range of detuning over which the high-spin ground state survives is larger than the hybridization $\sim40\,\mu$eV, consistent with the experiment\cite{dehollain2019nagaoka}. This can be reflected by the excited-state spectrum in Fig.~\ref{fig:detuningExcitedState}: the transition between low-spin and high-spin states occurs ``adiabatically'' between the ground states of each section. The Nagaoka gap is always much smaller than the level spacing, which is roughly reflected by the gap between the high-spin ground and excited states.

On the other hand, it is much easier to empty a site compared to doubly occupying one, in a hole-doped system (with three electrons on four sites): the detuning potential only has to compensate the kinetic energy instead of interaction energy to achieve the former. Thus, with a negative $dV$, the $E-dV$ slope flattens out rapidly, except for a small influence from the presence of hybridization. This means that increasing the site energy causes the emptying of the particular dot. For large enough $-dV$, the many-body system becomes an effective empty site plus three singly-occupied dots, or equivalently, a half-filled open-boundary array. Without the ``mobile'' hole in the ``half-filled'' system, the ground state becomes a low-spin state instead of the Nagaoka FM state. 

\begin{figure*}[!th]
\begin{center}
\includegraphics[width=2\columnwidth]{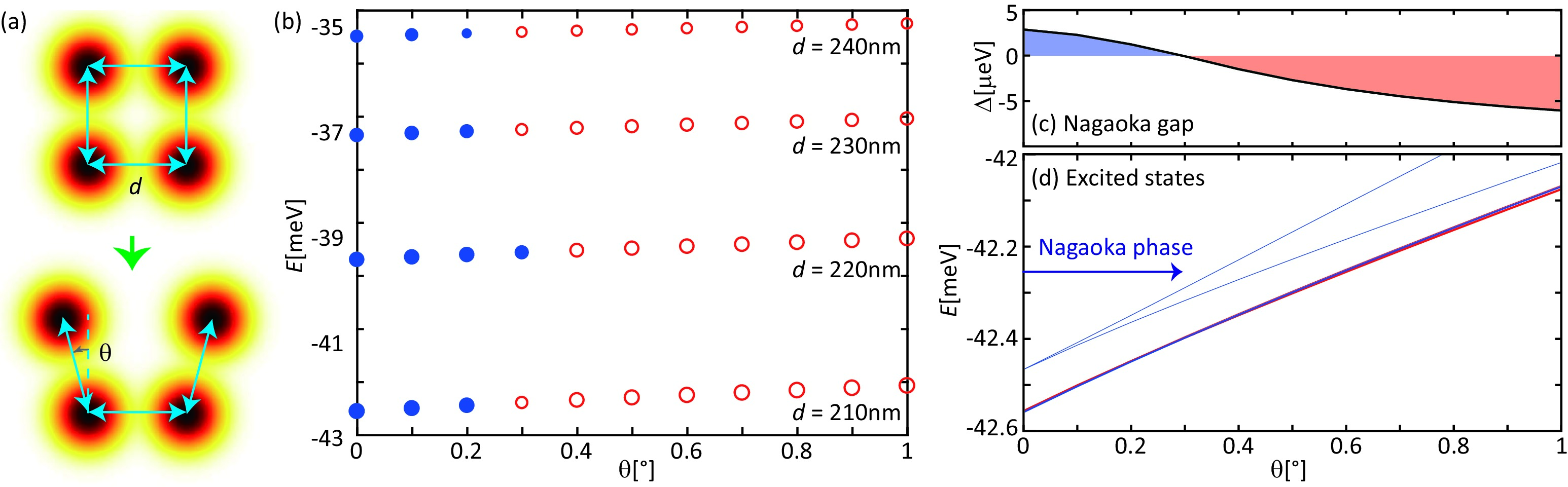}
\caption{\label{fig:angleGroundState} (a) Schematic illustrating the bond rotation in a $2\times2$ system. (b) The ground-state energy for the entire system as a function of the rotating angle $\theta$, calculated for various distances $d$. (c) The Nagaoka gap and (d) the first three excited-state energies in high-spin (blue) and low-spin (red) sectors for $d=210$\,nm. The arrow denotes the region of Nagaoka phase.
}
\end{center}
\end{figure*}

The effect of hybridization is made clear by the dot distance $d$ comparison in Fig.~\ref{fig:detuningGroundState}(b). With increasing distance, the slopes tend towards 0 for negative $dV$ and 1 for positive $dV$, since the increase in distance effectively suppresses any hybridization effects. Interestingly, although the Nagaoka gap decays rapidly for $d>210$\,nm in Fig.~\ref{fig:distDep}, it does not reflect the robustness against potential detuning. In fact, the range of $dV$ where the Nagaoka phase is retained is similar for $d=210$ and 220\,nm. Only after 220\,nm, the range starts to shrink. This is because the robustness of the Nagaoka phase not only depends on the absolute energy gap, but its relative strength compared to the effective tunneling $t$. The fact that $t$ drops by a factor of 2 from $d=210$\,nm to $d=220$\,nm compensates the reduction of the absolute Nagaoka gap.

\subsection{From a Plaquette to a Chain}

By increasing the distance between two of the dots in the plaquette, we can study the four-dot system under different topologies. The Nagaoka theorem applies to a 2D system with periodic boundary conditions. In contrast, a 1D open-boundary system must obey the Lieb-Mattis theorem, which restricts the ground-state solution to the lowest spin sector\cite{lieb1962lieb}. We can gradually change the topology, from a plaquette to a chain, by increasing the angle $\theta$ between two edges in the 2$\times$2 system, as shown in Fig.~\ref{fig:angleGroundState}(a). We again focus on the $d=210$\,nm system first. As shown in Fig.~\ref{fig:angleGroundState}(b), the ground state soon becomes a low-spin state for a rotation angle of $\sim$0.3$^\circ$ [also see Fig.~\ref{fig:angleGroundState}(c)]. The rapid increase of the ground-state energy indicates its sensitivity to the angle, or the topology. This sensitivity can be understood from the excited-state spectrum. The original plaquette has a $C_4$ rotational symmetry, leading to a rotational symmetric ground state. The first and second excited states correspond to the eigenstates of rotation with a factor of $e^{i\pi/2}$ and $e^{-i\pi/2}$, which are degenerate for $\theta=0$. Thus, the extent to which the system ceases to be 2D can be reflected by the energy splitting of these two excited states. As shown in Fig.~\ref{fig:angleGroundState}(d), these two lowest excited states soon separate from each other and the separation becomes comparable with the gap to the ground state for $\theta\sim 0.5^\circ$. This phenomenon indirectly reflects the fact that the system, including its ground state, becomes more like 1D in contrast to 2D, resulting in an $S=1/2$ instead of $S=3/2$ ground state.

Interestingly, the transition from high- to low-spin ground state occurs at very small angles, far before the system becomes 1D geometrically. As Mattis has pointed out, the Lieb-Mattis theorem holds only for a strictly 1D open-boundary system \cite{mattis2003eigenvalues}. That being said, there should be additional mechanisms accounting for the drop of Nagaoka ferromagnetism. The answer to this question might come from the intuition that Nagaoka ferromagnetism is a consequence of constructive interference between the paths that the hole can take through the plaquette, lowering the kinetic energy in the presence of $C_4$ rotational symmetry. This interference effect is quickly lost at even small values of $\theta$, with the broken rotational symmetry leading to unbalanced $x$- and $y$-direction hopping.

Alternatively, one can look at the above reasoning in terms of translational symmetry. Once the hopping between any neighboring sites is dramatically weakened, the system behaves more like an open-boundary chain describable by the Lieb-Mattis theorem. In this sense, the high- to low-spin transition is caused by unbalanced tunneling in the system, rather than geometry. In the experiment, the geometric modification of the system is achieved by tuning the gate potentials, which has a combined effect of increasing the potential barrier between the dots, as well as increasing their separation.

We also examine the transition for different distances $d$, as shown in Fig.~\ref{fig:angleGroundState}(b). Here we observe that the Nagaoka high-spin state is almost equally robust as a function of distance. This can be attributed to the fact that the intrinsic interaction and tunneling scales are almost unchanged when one rotates the two edges, especially for larger distances where the hybridization is negligible. In the former case, only the relative values of the tunneling strengths show up in the path interference, which depends on the rotation angle instead of the absolute tunneling strengths.


\section{Discussion and Conclusion}\label{sec:conclusion}
We described a theoretical, \emph{ab initio} analysis of a quantum dot plaquette system, in which we obtained quantitative agreement with the recent experimental study of the emergence of quantum magnetism through the Nagaoka mechanism. Our work provides theoretical support for the experimentally observed robustness of the Nagaoka state against perturbations such as distance between the dots and potential detuning. Interestingly, one can also find good agreement between experiments and a single-band extended Hubbard model by properly choosing model parameters\cite{dehollain2019nagaoka}. The effective single band should be understood as being comprised of a linear superposition of single-particle electronic orbitals determined by strong inter-orbital interactions. This phenomenological approach, however, has very limited predictive power as it fully relies on fitting parameters to experimental measurements. Our analysis demonstrates that \emph{ab initio} calculations are possible for experimentally relevant systems and can be used to study phenomena beyond the single-band model\cite{nelson2004odd,wu2008orbital,kim2009phase}. Even for the quantitative modeling on a single-band level, we expect the ``bottom-up'' approach to be more accurate than fitting to experimental data. Current experiments can only provide limited information about the excited states and gap sizes, even with the state-of-the-art experimental techniques, and do not allow to determine all parameters of the effective model. With a practical down-folding to the fewer-orbital models, one can further extend the calculation approximately to much larger quantum-dot systems.

Thus, with the focus on a tunable quantum-dot system, we have introduced the \emph{ab initio} exact diagonalization approach, which can be in general applied to different types of artificial quantum simulators. The computational complexity for the model parameter evaluation scales polynomially with the number of sites and orbitals. Calculating the expensive two-center integrals is most costly in the plaquette system. The next level of complexity for these calculations would consider multiple and inhomogeneous Gaussian decompositions, which are significant for stronger hybridized systems or higher-order corrections. These issues have been overcome in modern quantum chemistry using composite atomic basis. Through appropriate fitting using an extended Gaussian basis, we expect to solve these issues by the same means. In any case, the bottleneck of the \emph{ab initio} calculation comes from the bottom-level one-center and two-center integrals Eqs.~\eqref{eq:oneCenterIntegral} and \eqref{eq:twoCenterIntegral}. They have been shown to be efficiently accelerated using GPU-based programming, which can also be directly ported into our systems.

The evaluation of many-body model parameters through the \emph{ab initio} calculation has achieved the goal of precisely modeling an artificial electronic system. Although we here adopt the four-well system and the Nagaoka transition as an example of our approach, motivated by the recent quantum-dot experiment, we would like to emphasize that the \emph{ab initio} exact diagonalization approach can be applied to larger quantum-dot systems with necessary numerical improvements. Unlike the traditional mean-field approaches, a many-body numerical solver like exact diagonalization is always necessary to obtain the ground-state or excited-state wavefunctions. This step is relatively cheap in the current example, but scales up exponentially with the number of sites and electrons. To simulate a larger system, a proper separation of scales might be necessary. For example, if the electron occupation is large, the ``fully occupied'' low energy states may be treated by mean-field theory as a pseudo-potential, to limit the complexity to the bands near the Fermi level. Additionally, the efficiency of the modeling may be further increased employing other many-body numerical approaches including quantum Monte Carlo, density matrix renormalization group, embedding theory, and quantum cluster methods, depending on the purpose of calculation.

Focusing specifically on quantum dot simulators, the accessibility of multiple orbitals and precise treatment of electron interactions could enable a direct simulation of many-body states. Owing to the tunability and measurability of electronic configurations, the quantum dots have been shown to emulate artificial chemical molecules with dominant 2D geometry. For example, the four-dot system investigated in this work can be regarded as an H$_4$ molecule, which is a standard platform for testing correlated quantum chemistry methods. Hence the quantum-dot simulators can be used to find the many-body electron state in a Born-Oppenheimer assumption. 

Looking beyond quantum dot systems, this approach can be naturally extended to Rydberg atoms or cold molecules by replacing the Coulomb interaction $W(\rbf_1-\rbf_2)$ with the Lennard-Jones potential and making $V(\rbf)$ a standing-wave potential. The breaking of rotational symmetry in $V(\rbf)$ may cause more computational complexity, which can be overcome using some of the efficient integration implementations mentioned above. Moreover, the majority of the optical lattice studies concern bosons. The \emph{ab initio} exact diagonalization framework can be applied to bosonic systems by replacing the fermionic basis states represented by Slater determinants with bosonic product states represented by permanents. In general, this approach holds the promise to push the boundaries of predictability and quantitative accuracy in the ever-expanding zoo of quantum simulators that are being implemented.

\section*{Acknowledgements}
We thank for insightful discussions with S. Das Sarma and B. Wunsch. This work was supported at Harvard University by the Harvard-MIT Center for Ultracold Atoms, NSF Grant No.~DMR-1308435, the AFOSR-MURI Photonic Quantum Matter (Award No.~FA95501610323), and the DARPA DRINQS program (Award No.~D18AC00014). Y.W. was supported by the Postdoctoral Fellowship in Quantum Science of the Harvard-MPQ Center for Quantum Optics. L.M.K.V. thanks the NSF-funded MIT-Harvard Center for Ultracold Atoms for its hospitality. J.P.D., U.M., and L.M.K.V. acknowledge grants from the Netherlands Organisation for Scientific Research (FOM projectruimte and NWO Vici). M.R. thanks the Villum Foundation for support. This research used resources of the National Energy Research Scientific Computing Center (NERSC), a U.S. Department of Energy Office of Science User Facility operated under Contract No. DE-AC02-05CH11231.

\appendix
\section{Derivation of the Interacting Part of the Hamiltonian}\label{app:interactiongHam}
Restricting to the four-fermion terms, the second-quantized Hamiltonian can be generically expressed as\cite{sheng2005multiband}
\begin{widetext}
\begin{eqnarray}\label{eq:genericFourFermionTerms}
\Ham_{\rm int} = \frac12\sum_{i_1\alpha_1\sigma_1}\sum_{i_2\alpha_2\sigma_2}\sum_{j_1\beta_1\sigma^\prime_1}\sum_{j_2\beta_2\sigma^\prime_2} \mathcal{W}(j_1, \beta_1,\sigma^\prime_1; j_2, \beta_2,\sigma^\prime_2 |i_2, \alpha_2,\sigma_2; i_1, \alpha_1,\sigma_1) 
 c^\dagger_{j_1\beta_1\sigma^\prime_1}c^\dagger_{j_2\beta_2\sigma^\prime_2}  c_{i_2\alpha_2\sigma_2}c_{i_1\alpha_1\sigma_1} \,.
\end{eqnarray}
Substituting the wavefunctions into it, we obtain 
\begin{eqnarray}\label{eq:generalVterms}
    &&\mathcal{W}(j_1, \beta_1,\sigma^\prime_1; j_2, \beta_2,\sigma^\prime_2 |i_2, \alpha_2,\sigma_2; i_1, \alpha_1,\sigma_1) \nonumber\\
    &=&\sum_{{s_{z}}_1,s_{z2}}\iint d\rbf_1^dd\rbf_2^d W(|\rbf_1 -\rbf_2|)\Psi_{(j_1, \beta_1,\sigma^\prime_1),(j_2, \beta_2,\sigma^\prime_2)}(\rbf_1,\rbf_2)^*\Psi_{(i_1,\alpha_1,\sigma_1),(i_2,\alpha_2,\sigma_2)}(\rbf_1,\rbf_2)\nonumber\\
    &=& 
    \frac12\sum_{s_{z1},s_{z2}}\!\iint\! d\rbf_1^dd\rbf_2^d W(|\rbf_1 -\rbf_2|) \Big[\tilde\psi_{j_1 \beta_1\sigma^\prime_1}\!(\rbf_1)^*\tilde\psi_{j_2 \beta_2\sigma^\prime_2}\!(\rbf_2)^*-\tilde\psi_{j_2 \beta_2\sigma^\prime_2}\!(\rbf_1)^*\tilde\psi_{j_1 \beta_1\sigma^\prime_1}\!(\rbf_2)^*\Big]\nonumber\\
    &&\Big[\tilde\psi_{i_1\alpha_1\sigma_1}\!(\rbf_1)\tilde\psi_{i_2\alpha_2\sigma_2}\!(\rbf_2)-\tilde\psi_{i_2\alpha_2\sigma_2}\!(\rbf_1)\tilde\psi_{i_1\alpha_1\sigma_1}\!(\rbf_2)\Big]\nonumber\\
    &=& \iint d\rbf_1^dd\rbf_2^d W(|\rbf_1\! -\!\rbf_2|) \Big[\tilde\psi_{j_1 \beta_1\sigma^\prime_1}\!(\rbf_1)^*\tilde\psi_{j_2 \beta_2\sigma^\prime_2}\!(\rbf_2)^*\tilde\psi_{i_1\alpha_1\sigma_1}\!(\rbf_1)\tilde\psi_{i_2\alpha_2\sigma_2}\!(\rbf_2)\delta_{\sigma_1^\prime\sigma_1}\delta_{\sigma_2^\prime\sigma_2} \nonumber\\
    &&-\tilde\psi_{j_1 \beta_1\sigma^\prime_1}\!(\rbf_1)^*\tilde\psi_{j_2 \beta_2\sigma^\prime_2}\!(\rbf_2)^*\tilde\psi_{i_2\alpha_2\sigma_2}(\rbf_1)\tilde\psi_{i_1\alpha_1\sigma_1}\!(\rbf_2)\delta_{\sigma_1^\prime\sigma_2}\delta_{\sigma_2^\prime\sigma_1}\Big].
\end{eqnarray}
\end{widetext}

 Using the simplification mentioned in the main text, the interaction terms can be categorized into  $\sigma_1 = \sigma_1^\prime$ and $\sigma_1=\sigma_2^\prime$ parts. Denoting these two parts as $U$ and $J$, we obtain 
\begin{eqnarray}\label{eq:generalUJ}
\Ham_{\rm int} \!&=&\! \sum_{i,j\atop\sigma,\sigma^\prime}\!\sum_{\alpha_1,\alpha_2\atop\beta_1,\beta_2}\frac{U_{ij}^{\sigma\sigma^\prime}(\beta_1, \! \beta_2 |\alpha_2,\! \alpha_1) }{2}
 c^\dagger_{i\beta_1\sigma}c^\dagger_{j\beta_2\sigma^\prime}  c_{j\alpha_2\sigma^\prime}c_{i\alpha_1\sigma} \nonumber\\
 &&\!+ \!\sum_{i,j\atop\sigma,\sigma^\prime}\!\sum_{\alpha_1,\alpha_2\atop\beta_1,\beta_2} \!\frac{J_{ij}^{\sigma\sigma^\prime}\!( \beta_1,\! \beta_2 |\alpha_2,\! \alpha_1) }{2}
 c^\dagger_{i\beta_1\sigma^\prime}c^\dagger_{j\beta_2\sigma}  c_{j\alpha_2\sigma^\prime}c_{i\alpha_1\sigma}\,.
\end{eqnarray}
Note the $U$ and $J$ terms are not completely independent, since $U_{ij}^{\sigma\sigma}\equiv J_{ij}^{\sigma\sigma}$. Additionally, we also have the permutation symmetry
\begin{eqnarray}\label{eq:generalOnSiteTermsForm}
U_{ij}^{\sigma\sigma^\prime}(\beta_1,  \beta_2 |\alpha_2, \alpha_1) = U_{ji}^{\sigma^\prime\sigma}(\beta_2,  \beta_1 |\alpha_1, \alpha_2) \nonumber\\
J_{ij}^{\sigma\sigma^\prime}(\beta_1,  \beta_2 |\alpha_2, \alpha_1) = J_{ji}^{\sigma^\prime\sigma}(\beta_2,  \beta_1 |\alpha_1, \alpha_2) \,.
\end{eqnarray}

For electrons on a single lattice site, the generic form in Eq.~\eqref{eq:generalUJ} reduces to
\begin{eqnarray}\label{eq:fullOnSiteHamiltonian}
\Ham^{\rm (OS)}_i&=&\frac12\sum_{\alpha\sigma} U_{\alpha} n_{\alpha\bar{\sigma}}n_{\alpha\sigma} + \frac12\sum_{\alpha_1\neq\alpha_2}\sum_{\sigma} U^\prime_{\alpha_1 \alpha_2} n_{\alpha_2\sigma}n_{\alpha_1\sigma} \nonumber\\
 &&+  \frac12\sum_{\alpha_1\neq\alpha_2}\sum_{\sigma} \bar{U}^\prime_{\alpha_1 \alpha_2} n_{\alpha_2\bar{\sigma}}n_{\alpha_1\sigma}\nonumber\\
 &&  +  \frac12\sum_{\alpha_1\neq\alpha_2}\sum_{\sigma} J_{\alpha_1 \alpha_2}
 c^\dagger_{\alpha_2\sigma}c^\dagger_{\alpha_1\sigma}  c_{\alpha_2\sigma}c_{\alpha_1\sigma} \nonumber\\
 &&+  \frac12\sum_{\alpha_1\neq\alpha_2}\sum_{\sigma} \bar{J}_{\alpha_1 \alpha_2}
 c^\dagger_{\alpha_2\sigma}c^\dagger_{\alpha_1\bar{\sigma}}  c_{\alpha_2\bar{\sigma}}c_{\alpha_1\sigma}\,.
\end{eqnarray}
Note that the spin-parallel Hund term  $J_{\alpha_1 \alpha_2}$ is the same as the spin-parallel Hubbard term $U^\prime_{\alpha_1 \alpha_2}$ with a sign flip.

Therefore, the on-site Hubbard interaction in Eq.~\eqref{eq:generalOnSiteTermsForm} is
\begin{equation}
	\begin{aligned}
	U_{\alpha}\! &=\mathcal{W}(i, \alpha,\sigma; i, \alpha,\bar{\sigma} |i, \alpha,\bar{\sigma}; i, \alpha,\sigma)\\
    &=\! \iint\! d\rbf_1^dd\rbf_2^d W(|\rbf_1\! -\!\rbf_2|) |\tilde\psi_{i\alpha\sigma}(\rbf_1)|^2|\tilde\psi_{i\alpha\bar{\sigma}}(\rbf_2)|^2\,.
	\end{aligned}
\end{equation}
The Hubbard interaction is dominant among the interaction terms due to the maximal overlap of wavefunctions. The remaining terms in a single-well interaction are all the inter-orbital interactions. The spin-parallel interaction is
\begin{eqnarray}\label{eq:parallelIntegral}
    &&U^\prime_{\alpha_1 \alpha_2} - J_{\alpha_1 \alpha_2} \nonumber\\
    &=&\mathcal{W}(i, \alpha_1,\sigma; i, \alpha_2,\sigma |i, \alpha_2,\sigma; i, \alpha_1,\sigma)\nonumber\\
    &=& \iint d\rbf_1^dd\rbf_2^d W(|\rbf_1\! -\!\rbf_2|)\Big[|\tilde\psi_{i \alpha_1\sigma}(\rbf_1)|^2|\tilde\psi_{i \alpha_2\sigma}(\rbf_2)|^2 \nonumber\\
 &&-\tilde\psi_{i \alpha_1\sigma}\!(\rbf_1)^*\tilde\psi_{i \alpha_2\sigma}\!(\rbf_2)^*\tilde\psi_{i\alpha_2\sigma}(\rbf_1)\tilde\psi_{i\alpha_1\sigma}\!(\rbf_2)\Big],
\end{eqnarray}
while spin-anti-parallel interaction is
\begin{equation}\label{eq:antiParallelIntegral}
	\begin{aligned}
    \bar{U}^\prime_{\alpha_1 \alpha_2}\! &=\mathcal{W}(i, \alpha_1,\sigma; i, \alpha_2,\bar{\sigma} |i, \alpha_2,\bar{\sigma}; i, \alpha_1,\sigma)\\
    &=\! \iint\! d\rbf_1^dd\rbf_2^d W(|\rbf_1\! -\!\rbf_2|)|\tilde\psi_{i \alpha_1\sigma}(\rbf_1)|^2|\tilde\psi_{i \alpha_2\bar{\sigma}}(\rbf_2)|\,.
\end{aligned}
\end{equation}
Given that the two-body interaction $W(|\rbf_1 -\rbf_2|)$ (typically Coulomb) does not involve spin degrees of freedom, the first term of Eq.~\eqref{eq:parallelIntegral} is equal to the anti-parallel spin contribution in Eq.~\eqref{eq:antiParallelIntegral}. Naturally, one can split the entire parallel spin interactions in Eq.~\eqref{eq:parallelIntegral} into charge and Hund's part by assuming $U^\prime_{\alpha_1 \alpha_2} = \bar{U}^\prime_{\alpha_1 \alpha_2}$. This partition also guarantees the equivalence of the two exchange coefficients
\begin{eqnarray}
    \bar{J}_{\alpha_1 \alpha_2} &=&\mathcal{W}(i, \alpha_2,\sigma; i, \alpha_1,\bar{\sigma} |i, \alpha_2,\bar{\sigma}; i, \alpha_1,\sigma)\nonumber\\
    &=& \iint d\rbf_1^dd\rbf_2^d W(|\rbf_1\! -\!\rbf_2|)\tilde\psi_{i \alpha_2\sigma}(\rbf_1)^*\tilde\psi_{i\alpha_1\bar{\sigma}}(\rbf_2)^*\!\nonumber\\
&&\tilde\psi_{i\alpha_1\sigma}(\rbf_1)\tilde\psi_{i\alpha_2\bar{\sigma}}(\rbf_2)\nonumber\\
&=& J_{\alpha_1 \alpha_2}\,.
\end{eqnarray} 
Therefore, we obtain the on-site interacting Hamiltonian Eq.~\eqref{eq:multipletHam} in the main text.

Then following Eq.~\eqref{eq:longrangeHam} in the main text, we can evaluate the interaction parameters in the long-range part of Hamiltonian $\Ham_{ij}^{\rm(LR)}$. 
Similar to the on-site terms, the ``off-diagonal'' terms of $V_{\alpha\beta}$ and $V^\prime_{\alpha\beta}$ are absorbed by the corresponding exchange terms for parallel spins. Therefore, we can write expressions for each of the relevant long-range terms
\begin{eqnarray}
    V_{\alpha\beta} &=& \iint d\rbf_1^dd\rbf_2^d \,W(|\rbf_1\! -\!\rbf_2|)\,|\tilde\psi_{i \alpha\sigma}(\rbf_1)|^2\,|\tilde\psi_{j \beta\sigma}(\rbf_2)|^2\nonumber\\
    K_{\alpha\beta} &=&\iint d\rbf_1^dd\rbf_2^d\, W(|\rbf_1\! -\!\rbf_2|)\,\tilde\psi_{j \beta\sigma}\!(\rbf_1)^*\tilde \psi_{i\alpha\sigma^\prime}(\rbf_2)^*\nonumber\\
    &&\tilde\psi_{i\alpha\sigma}(\rbf_1)\tilde\psi_{j\beta\sigma^\prime}(\rbf_2)\nonumber\\
    V^\prime_{\alpha\beta} &=&\iint d\rbf_1^dd\rbf_2^d\, W(|\rbf_1\! -\!\rbf_2|)\,\tilde\psi_{i \beta\sigma}(\rbf_1)^*\tilde\psi_{j\alpha\sigma^\prime}(\rbf_2)^*\nonumber\\
    &&\tilde\psi_{i\alpha\sigma}(\rbf_1)\tilde\psi_{j\beta\sigma^\prime}(\rbf_2)\nonumber\\
    K^\prime_{\alpha\beta} &=& \iint d\rbf_1^dd\rbf_2^d\, W(|\rbf_1\! -\!\rbf_2|)\,\tilde\psi_{j \alpha\sigma}(\rbf_1)^*\tilde\psi_{i\beta\sigma^\prime}(\rbf_2)^*\nonumber\\
    &&\tilde\psi_{i\alpha\sigma}(\rbf_1)\tilde\psi_{j\beta\sigma^\prime}(\rbf_2)\,,
\end{eqnarray}
and transform them to the original basis, resulting in Eq.~\eqref{eq:longrangeParams} in the main text.

\section{Evaluation of the Integrals}\label{app:integral}
The algebraic representations of the model parameters Eqs.~\eqref{eq:onsiteParams} and \eqref{eq:longrangeParams} concentrate all integration calculations in the evaluation of $\Xi_{\mu_1\nu_1\atop\mu_2\nu_2}$ in the single-well basis. This evaluation is not trivial, since the direct expression Eq.~\eqref{eq:expXi} contains a $2\times d$-dimensional integral with singularities, which cannot be computed efficiently even with supercomputers\cite{helgaker2014molecular}. However, taking advantage of the rotational symmetry of the quantum well, the calculation can be significantly simplified.

Let us first look at the dominant part -- the one-center integral, where all four wavefunctions are centered in the same well. Taking advantage of the rotational invariance of $W(|\rbf_1\! -\!\rbf_2|)$, one can simplify the integral through the Wigner-Eckart theorem. Specifically, for a Coulomb-type interaction, we have the Laplacian expansion
\begin{equation}
    \frac{1}{|\rbf_1-\rbf_2|} = \frac{1}{r_>}\sum_{l \geq 0} \left( \frac{r_<}{r_>}\right)^l P_{l}\big(\cos{(\theta_1-\theta_2)}\big),
\end{equation}
in which $P_{l}(x)$ is the Legendre polynomial. With $W(|\rbf_1\! -\!\rbf_2|)=e^2/4\pi\epsilon|\rbf_1\! -\!\rbf_2|$, we can simplify the one-center integral to a sequence of two-dimensional integrals as
\begin{eqnarray}\label{eq:oneCenterIntegral}
    \Gamma_{\mu_1\nu_1\atop\mu_2\nu_2}^{(l)}&= & \int_0^{+\infty} \int_0^{r_1} \,dr_2dr_1 \frac1{r_1} \left( \frac{r_2}{r_1}\right)^l  \chi_{\mu_1}(r_1)\chi_{\nu_1}(r_1)\nonumber\\&&\chi_{\mu_2}(r_2)\chi_{\nu_2}(r_2) \nonumber\\
    \Theta_{\mu_1\nu_1\atop\mu_2\nu_2}^{(l)} &=& \iint_0^{2\pi}\, d\phi_1d\phi_2 P_l\big(\cos{(\phi_1\!-\!\phi_2)}\big)\varphi_{\mu_1}(\phi_1)\varphi_{\mu_2}(\phi_2)\nonumber\\&&\varphi_{\nu_1}(\phi_1)\varphi_{\nu_2}(\phi_2),
\end{eqnarray}
then Eq.~\eqref{eq:expXi} is expanded as 
\begin{eqnarray}
    \Xi_{\mu_1\nu_1\atop\mu_2\nu_2}= \frac{e^2}{4\pi\epsilon}\sum_{l = 0}^{\infty}\left(\Gamma_{\mu_1\nu_1\atop\mu_2\nu_2}^{(l)} + \Gamma_{\mu_2\nu_2\atop\mu_1\nu_1}^{(l)}\right)\Theta_{\mu_1\nu_1\atop\mu_2\nu_2}^{(l)}\,.
\end{eqnarray}
Note, that $\Theta_{\mu_1\nu_1\atop\mu_2\nu_2}^{(l)}$ is symmetric under exchange of 1 and 2 indices, while $\Xi_{\mu_1\nu_1\atop\mu_2\nu_2}$ is usually not symmetric except in special cases where $\{\mu_1,\nu_1\} = \{\mu_2,\nu_2\}$. The integral decays rapidly with the increase of $l$. With fine enough spatial grids and angular momentum truncation, the one-center integral can be evaluated up to machine precision.

In contrast, the two-center integral involves more computational complexity. Here, rotational symmetry is not maintained; therefore, there is no direct separation of variables. However, we know that the ground state and the norm of low-lying excited-state wavefunctions can be well estimated by different Gaussian functions. This provides a way to estimate the density-density correlation among the two-center integrals. If the density distribution is written as 
\begin{equation}
    n(\rbf; \Rbf, \sigma) = \frac1{2\pi\sigma^2}e^{-\frac{(\rbf-\Rbf)^2}{2\sigma^2}}\,,
\end{equation}
the two-center integral can be decomposed in the center-of-mass frame
\begin{eqnarray}\label{eq:twoCenterIntegral}
    &&\iint  n(\rbf_1; \Rbf_1,\sigma_1) \frac1{|\rbf_1-\rbf_2|}n(\rbf_2; \Rbf_2,\sigma_2) d\rbf_1^2 d\rbf_2^2\nonumber\\
    &=&\frac1{4\pi^2(\sigma_1^2+\sigma_2^2)^2}\iint e^{-\frac{(\bar{\rbf}-\bar{\Rbf})^2}{2(\sigma_1^2+\sigma_2^2)}}\frac1{|\Delta\rbf|}e^{-\frac{(\Delta{\rbf}-\Delta{\Rbf})^2}{2(\sigma_1^2+\sigma_2^2)}}d\bar{\rbf}^2 d\Delta\rbf^2\nonumber\\
    &=&\frac1{2\pi(\sigma_1^2+\sigma_2^2)}\iint\frac1{|\Delta\rbf|}e^{-\frac{(\Delta{\rbf}-\Delta{\Rbf})^2}{2(\sigma_1^2+\sigma_2^2)}} d\Delta\rbf^2\,.
\end{eqnarray}
Now, the integral is reduced to a two-dimensional integral in the reduced coordinates $\Delta \rbf$, which can be solved by using the Riemann integral or the Laplacian expansion as mentioned above. Note that the Gaussian integral provides only an estimation of the realistic two-center interaction. A more precise treatment involves the decomposition of multiple Gaussian bases and its derivatives\cite{mcmurchie1978one}, which forms the foundation of electronic structure theory and is beyond the scope of this work.
\bibliography{paper}

\begin{thebibliography}{110}
\expandafter\ifx\csname natexlab\endcsname\relax\def\natexlab#1{#1}\fi
\expandafter\ifx\csname bibnamefont\endcsname\relax
  \def\bibnamefont#1{#1}\fi
\expandafter\ifx\csname bibfnamefont\endcsname\relax
  \def\bibfnamefont#1{#1}\fi
\expandafter\ifx\csname citenamefont\endcsname\relax
  \def\citenamefont#1{#1}\fi
\expandafter\ifx\csname url\endcsname\relax
  \def\url#1{\texttt{#1}}\fi
\expandafter\ifx\csname urlprefix\endcsname\relax\def\urlprefix{URL }\fi
\providecommand{\bibinfo}[2]{#2}
\providecommand{\eprint}[2][]{\url{#2}}

\bibitem[{\citenamefont{Bednorz and M{\"u}ller}(1986)}]{bednorz1986possible}
\bibinfo{author}{\bibfnamefont{J.~G.} \bibnamefont{Bednorz}} \bibnamefont{and}
  \bibinfo{author}{\bibfnamefont{K.~A.} \bibnamefont{M{\"u}ller}},
  \bibinfo{journal}{Z. Angew. Phys. B Condens. Mat.}
  \textbf{\bibinfo{volume}{64}}, \bibinfo{pages}{189} (\bibinfo{year}{1986}).

\bibitem[{\citenamefont{Fazekas}(1999)}]{fazekas1999lecture}
\bibinfo{author}{\bibfnamefont{P.}~\bibnamefont{Fazekas}},
  \emph{\bibinfo{title}{Lecture notes on electron correlation and magnetism}},
  vol.~\bibinfo{volume}{5} (\bibinfo{publisher}{World Scientific, Singapore},
  \bibinfo{year}{1999}).

\bibitem[{\citenamefont{Stormer and Tsui}(1983)}]{stormer1983quantized}
\bibinfo{author}{\bibfnamefont{H.}~\bibnamefont{Stormer}} \bibnamefont{and}
  \bibinfo{author}{\bibfnamefont{D.}~\bibnamefont{Tsui}},
  \bibinfo{journal}{Science} \textbf{\bibinfo{volume}{220}},
  \bibinfo{pages}{1241} (\bibinfo{year}{1983}).

\bibitem[{\citenamefont{Dagotto}(1994)}]{dagotto1994correlated}
\bibinfo{author}{\bibfnamefont{E.}~\bibnamefont{Dagotto}},
  \bibinfo{journal}{Rev. Mod. Phys.} \textbf{\bibinfo{volume}{66}},
  \bibinfo{pages}{763} (\bibinfo{year}{1994}).

\bibitem[{\citenamefont{Lin}(1990)}]{lin1990exact}
\bibinfo{author}{\bibfnamefont{H.}~\bibnamefont{Lin}}, \bibinfo{journal}{Phys.
  Rev. B} \textbf{\bibinfo{volume}{42}}, \bibinfo{pages}{6561}
  (\bibinfo{year}{1990}).

\bibitem[{\citenamefont{Foulkes et~al.}(2001)\citenamefont{Foulkes, Mitas,
  Needs, and Rajagopal}}]{foulkes2001quantum}
\bibinfo{author}{\bibfnamefont{W.}~\bibnamefont{Foulkes}},
  \bibinfo{author}{\bibfnamefont{L.}~\bibnamefont{Mitas}},
  \bibinfo{author}{\bibfnamefont{R.}~\bibnamefont{Needs}}, \bibnamefont{and}
  \bibinfo{author}{\bibfnamefont{G.}~\bibnamefont{Rajagopal}},
  \bibinfo{journal}{Rev. Mod. Phys.} \textbf{\bibinfo{volume}{73}},
  \bibinfo{pages}{33} (\bibinfo{year}{2001}).

\bibitem[{\citenamefont{Schollw{\"o}ck}(2005)}]{schollwock2005density}
\bibinfo{author}{\bibfnamefont{U.}~\bibnamefont{Schollw{\"o}ck}},
  \bibinfo{journal}{Rev. Mod. Phys.} \textbf{\bibinfo{volume}{77}},
  \bibinfo{pages}{259} (\bibinfo{year}{2005}).

\bibitem[{\citenamefont{Bloch et~al.}(2008)\citenamefont{Bloch, Dalibard, and
  Zwerger}}]{bloch2008many}
\bibinfo{author}{\bibfnamefont{I.}~\bibnamefont{Bloch}},
  \bibinfo{author}{\bibfnamefont{J.}~\bibnamefont{Dalibard}}, \bibnamefont{and}
  \bibinfo{author}{\bibfnamefont{W.}~\bibnamefont{Zwerger}},
  \bibinfo{journal}{Rev. Mod. Phys.} \textbf{\bibinfo{volume}{80}},
  \bibinfo{pages}{885} (\bibinfo{year}{2008}).

\bibitem[{\citenamefont{Lewenstein et~al.}(2012)\citenamefont{Lewenstein,
  Sanpera, and Ahufinger}}]{lewenstein2012ultracold}
\bibinfo{author}{\bibfnamefont{M.}~\bibnamefont{Lewenstein}},
  \bibinfo{author}{\bibfnamefont{A.}~\bibnamefont{Sanpera}}, \bibnamefont{and}
  \bibinfo{author}{\bibfnamefont{V.}~\bibnamefont{Ahufinger}},
  \emph{\bibinfo{title}{Ultracold Atoms in Optical Lattices: Simulating quantum
  many-body systems}} (\bibinfo{publisher}{Oxford University Press, Oxford,
  UK}, \bibinfo{year}{2012}).

\bibitem[{\citenamefont{Gross and Bloch}(2017)}]{gross2017quantum}
\bibinfo{author}{\bibfnamefont{C.}~\bibnamefont{Gross}} \bibnamefont{and}
  \bibinfo{author}{\bibfnamefont{I.}~\bibnamefont{Bloch}},
  \bibinfo{journal}{Science} \textbf{\bibinfo{volume}{357}},
  \bibinfo{pages}{995} (\bibinfo{year}{2017}).

\bibitem[{\citenamefont{Parsons et~al.}(2015)\citenamefont{Parsons, Huber,
  Mazurenko, Chiu, Setiawan, Wooley-Brown, Blatt, and
  Greiner}}]{parsons2015site}
\bibinfo{author}{\bibfnamefont{M.~F.} \bibnamefont{Parsons}},
  \bibinfo{author}{\bibfnamefont{F.}~\bibnamefont{Huber}},
  \bibinfo{author}{\bibfnamefont{A.}~\bibnamefont{Mazurenko}},
  \bibinfo{author}{\bibfnamefont{C.~S.} \bibnamefont{Chiu}},
  \bibinfo{author}{\bibfnamefont{W.}~\bibnamefont{Setiawan}},
  \bibinfo{author}{\bibfnamefont{K.}~\bibnamefont{Wooley-Brown}},
  \bibinfo{author}{\bibfnamefont{S.}~\bibnamefont{Blatt}}, \bibnamefont{and}
  \bibinfo{author}{\bibfnamefont{M.}~\bibnamefont{Greiner}},
  \bibinfo{journal}{Phys. Rev. Lett.} \textbf{\bibinfo{volume}{114}},
  \bibinfo{pages}{213002} (\bibinfo{year}{2015}).

\bibitem[{\citenamefont{Greif et~al.}(2016)\citenamefont{Greif, Parsons,
  Mazurenko, Chiu, Blatt, Huber, Ji, and Greiner}}]{greif2016site}
\bibinfo{author}{\bibfnamefont{D.}~\bibnamefont{Greif}},
  \bibinfo{author}{\bibfnamefont{M.~F.} \bibnamefont{Parsons}},
  \bibinfo{author}{\bibfnamefont{A.}~\bibnamefont{Mazurenko}},
  \bibinfo{author}{\bibfnamefont{C.~S.} \bibnamefont{Chiu}},
  \bibinfo{author}{\bibfnamefont{S.}~\bibnamefont{Blatt}},
  \bibinfo{author}{\bibfnamefont{F.}~\bibnamefont{Huber}},
  \bibinfo{author}{\bibfnamefont{G.}~\bibnamefont{Ji}}, \bibnamefont{and}
  \bibinfo{author}{\bibfnamefont{M.}~\bibnamefont{Greiner}},
  \bibinfo{journal}{Science} \textbf{\bibinfo{volume}{351}},
  \bibinfo{pages}{953} (\bibinfo{year}{2016}).

\bibitem[{\citenamefont{Parsons et~al.}(2016)\citenamefont{Parsons, Mazurenko,
  Chiu, Ji, Greif, and Greiner}}]{parsons2016site}
\bibinfo{author}{\bibfnamefont{M.~F.} \bibnamefont{Parsons}},
  \bibinfo{author}{\bibfnamefont{A.}~\bibnamefont{Mazurenko}},
  \bibinfo{author}{\bibfnamefont{C.~S.} \bibnamefont{Chiu}},
  \bibinfo{author}{\bibfnamefont{G.}~\bibnamefont{Ji}},
  \bibinfo{author}{\bibfnamefont{D.}~\bibnamefont{Greif}}, \bibnamefont{and}
  \bibinfo{author}{\bibfnamefont{M.}~\bibnamefont{Greiner}},
  \bibinfo{journal}{Science} \textbf{\bibinfo{volume}{353}},
  \bibinfo{pages}{1253} (\bibinfo{year}{2016}).

\bibitem[{\citenamefont{Boll et~al.}(2016)\citenamefont{Boll, Hilker, Salomon,
  Omran, Nespolo, Pollet, Bloch, and Gross}}]{boll2016spin}
\bibinfo{author}{\bibfnamefont{M.}~\bibnamefont{Boll}},
  \bibinfo{author}{\bibfnamefont{T.~A.} \bibnamefont{Hilker}},
  \bibinfo{author}{\bibfnamefont{G.}~\bibnamefont{Salomon}},
  \bibinfo{author}{\bibfnamefont{A.}~\bibnamefont{Omran}},
  \bibinfo{author}{\bibfnamefont{J.}~\bibnamefont{Nespolo}},
  \bibinfo{author}{\bibfnamefont{L.}~\bibnamefont{Pollet}},
  \bibinfo{author}{\bibfnamefont{I.}~\bibnamefont{Bloch}}, \bibnamefont{and}
  \bibinfo{author}{\bibfnamefont{C.}~\bibnamefont{Gross}},
  \bibinfo{journal}{Science} \textbf{\bibinfo{volume}{353}},
  \bibinfo{pages}{1257} (\bibinfo{year}{2016}).

\bibitem[{\citenamefont{Brown et~al.}(2019)\citenamefont{Brown, Mitra,
  Guardado-Sanchez, Nourafkan, Reymbaut, H{\'e}bert, Bergeron, Tremblay,
  Kokalj, Huse et~al.}}]{brown2019bad}
\bibinfo{author}{\bibfnamefont{P.~T.} \bibnamefont{Brown}},
  \bibinfo{author}{\bibfnamefont{D.}~\bibnamefont{Mitra}},
  \bibinfo{author}{\bibfnamefont{E.}~\bibnamefont{Guardado-Sanchez}},
  \bibinfo{author}{\bibfnamefont{R.}~\bibnamefont{Nourafkan}},
  \bibinfo{author}{\bibfnamefont{A.}~\bibnamefont{Reymbaut}},
  \bibinfo{author}{\bibfnamefont{C.-D.} \bibnamefont{H{\'e}bert}},
  \bibinfo{author}{\bibfnamefont{S.}~\bibnamefont{Bergeron}},
  \bibinfo{author}{\bibfnamefont{A.-M.} \bibnamefont{Tremblay}},
  \bibinfo{author}{\bibfnamefont{J.}~\bibnamefont{Kokalj}},
  \bibinfo{author}{\bibfnamefont{D.~A.} \bibnamefont{Huse}},
  \bibnamefont{et~al.}, \bibinfo{journal}{Science}
  \textbf{\bibinfo{volume}{363}}, \bibinfo{pages}{379} (\bibinfo{year}{2019}).

\bibitem[{\citenamefont{Nichols et~al.}(2019)\citenamefont{Nichols, Cheuk,
  Okan, Hartke, Mendez, Senthil, Khatami, Zhang, and
  Zwierlein}}]{nichols2019spin}
\bibinfo{author}{\bibfnamefont{M.~A.} \bibnamefont{Nichols}},
  \bibinfo{author}{\bibfnamefont{L.~W.} \bibnamefont{Cheuk}},
  \bibinfo{author}{\bibfnamefont{M.}~\bibnamefont{Okan}},
  \bibinfo{author}{\bibfnamefont{T.~R.} \bibnamefont{Hartke}},
  \bibinfo{author}{\bibfnamefont{E.}~\bibnamefont{Mendez}},
  \bibinfo{author}{\bibfnamefont{T.}~\bibnamefont{Senthil}},
  \bibinfo{author}{\bibfnamefont{E.}~\bibnamefont{Khatami}},
  \bibinfo{author}{\bibfnamefont{H.}~\bibnamefont{Zhang}}, \bibnamefont{and}
  \bibinfo{author}{\bibfnamefont{M.~W.} \bibnamefont{Zwierlein}},
  \bibinfo{journal}{Science} \textbf{\bibinfo{volume}{363}},
  \bibinfo{pages}{383} (\bibinfo{year}{2019}).

\bibitem[{\citenamefont{Van~der Wiel et~al.}(2002)\citenamefont{Van~der Wiel,
  De~Franceschi, Elzerman, Fujisawa, Tarucha, and
  Kouwenhoven}}]{van2002electron}
\bibinfo{author}{\bibfnamefont{W.~G.} \bibnamefont{Van~der Wiel}},
  \bibinfo{author}{\bibfnamefont{S.}~\bibnamefont{De~Franceschi}},
  \bibinfo{author}{\bibfnamefont{J.~M.} \bibnamefont{Elzerman}},
  \bibinfo{author}{\bibfnamefont{T.}~\bibnamefont{Fujisawa}},
  \bibinfo{author}{\bibfnamefont{S.}~\bibnamefont{Tarucha}}, \bibnamefont{and}
  \bibinfo{author}{\bibfnamefont{L.~P.} \bibnamefont{Kouwenhoven}},
  \bibinfo{journal}{Rev. Mod. Phys.} \textbf{\bibinfo{volume}{75}},
  \bibinfo{pages}{1} (\bibinfo{year}{2002}).

\bibitem[{\citenamefont{Hanson et~al.}(2007)\citenamefont{Hanson, Kouwenhoven,
  Petta, Tarucha, and Vandersypen}}]{hanson2007spins}
\bibinfo{author}{\bibfnamefont{R.}~\bibnamefont{Hanson}},
  \bibinfo{author}{\bibfnamefont{L.~P.} \bibnamefont{Kouwenhoven}},
  \bibinfo{author}{\bibfnamefont{J.~R.} \bibnamefont{Petta}},
  \bibinfo{author}{\bibfnamefont{S.}~\bibnamefont{Tarucha}}, \bibnamefont{and}
  \bibinfo{author}{\bibfnamefont{L.~M.} \bibnamefont{Vandersypen}},
  \bibinfo{journal}{Rev. Mod. Phys.} \textbf{\bibinfo{volume}{79}},
  \bibinfo{pages}{1217} (\bibinfo{year}{2007}).

\bibitem[{\citenamefont{Nielsen and Bhatt}(2007)}]{nielsen2007nanoscale}
\bibinfo{author}{\bibfnamefont{E.}~\bibnamefont{Nielsen}} \bibnamefont{and}
  \bibinfo{author}{\bibfnamefont{R.}~\bibnamefont{Bhatt}},
  \bibinfo{journal}{Phys. Rev. B} \textbf{\bibinfo{volume}{76}},
  \bibinfo{pages}{161202} (\bibinfo{year}{2007}).

\bibitem[{\citenamefont{Oguri et~al.}(2007)\citenamefont{Oguri, Nisikawa,
  Tanaka, and Numata}}]{oguri2007kondo}
\bibinfo{author}{\bibfnamefont{A.}~\bibnamefont{Oguri}},
  \bibinfo{author}{\bibfnamefont{Y.}~\bibnamefont{Nisikawa}},
  \bibinfo{author}{\bibfnamefont{Y.}~\bibnamefont{Tanaka}}, \bibnamefont{and}
  \bibinfo{author}{\bibfnamefont{T.}~\bibnamefont{Numata}},
  \bibinfo{journal}{J. Magn. Magn. Mater.} \textbf{\bibinfo{volume}{310}},
  \bibinfo{pages}{1139} (\bibinfo{year}{2007}).

\bibitem[{\citenamefont{Thalineau et~al.}(2012)\citenamefont{Thalineau,
  Hermelin, Wieck, B{\"a}uerle, Saminadayar, and Meunier}}]{thalineau2012few}
\bibinfo{author}{\bibfnamefont{R.}~\bibnamefont{Thalineau}},
  \bibinfo{author}{\bibfnamefont{S.}~\bibnamefont{Hermelin}},
  \bibinfo{author}{\bibfnamefont{A.~D.} \bibnamefont{Wieck}},
  \bibinfo{author}{\bibfnamefont{C.}~\bibnamefont{B{\"a}uerle}},
  \bibinfo{author}{\bibfnamefont{L.}~\bibnamefont{Saminadayar}},
  \bibnamefont{and} \bibinfo{author}{\bibfnamefont{T.}~\bibnamefont{Meunier}},
  \bibinfo{journal}{Appl. Phys. Lett.} \textbf{\bibinfo{volume}{101}},
  \bibinfo{pages}{103102} (\bibinfo{year}{2012}).

\bibitem[{\citenamefont{Seo et~al.}(2013)\citenamefont{Seo, Choi, Lee, Kim,
  Chung, Sim, Umansky, and Mahalu}}]{seo2013charge}
\bibinfo{author}{\bibfnamefont{M.}~\bibnamefont{Seo}},
  \bibinfo{author}{\bibfnamefont{H.}~\bibnamefont{Choi}},
  \bibinfo{author}{\bibfnamefont{S.-Y.} \bibnamefont{Lee}},
  \bibinfo{author}{\bibfnamefont{N.}~\bibnamefont{Kim}},
  \bibinfo{author}{\bibfnamefont{Y.}~\bibnamefont{Chung}},
  \bibinfo{author}{\bibfnamefont{H.-S.} \bibnamefont{Sim}},
  \bibinfo{author}{\bibfnamefont{V.}~\bibnamefont{Umansky}}, \bibnamefont{and}
  \bibinfo{author}{\bibfnamefont{D.}~\bibnamefont{Mahalu}},
  \bibinfo{journal}{Phys. Rev. Lett.} \textbf{\bibinfo{volume}{110}},
  \bibinfo{pages}{046803} (\bibinfo{year}{2013}).

\bibitem[{\citenamefont{Hensgens et~al.}(2017)\citenamefont{Hensgens, Fujita,
  Janssen, Li, Van~Diepen, Reichl, Wegscheider, Sarma, and
  Vandersypen}}]{hensgens2017quantum}
\bibinfo{author}{\bibfnamefont{T.}~\bibnamefont{Hensgens}},
  \bibinfo{author}{\bibfnamefont{T.}~\bibnamefont{Fujita}},
  \bibinfo{author}{\bibfnamefont{L.}~\bibnamefont{Janssen}},
  \bibinfo{author}{\bibfnamefont{X.}~\bibnamefont{Li}},
  \bibinfo{author}{\bibfnamefont{C.}~\bibnamefont{Van~Diepen}},
  \bibinfo{author}{\bibfnamefont{C.}~\bibnamefont{Reichl}},
  \bibinfo{author}{\bibfnamefont{W.}~\bibnamefont{Wegscheider}},
  \bibinfo{author}{\bibfnamefont{S.~D.} \bibnamefont{Sarma}}, \bibnamefont{and}
  \bibinfo{author}{\bibfnamefont{L.~M.} \bibnamefont{Vandersypen}},
  \bibinfo{journal}{Nature} \textbf{\bibinfo{volume}{548}}, \bibinfo{pages}{70}
  (\bibinfo{year}{2017}).

\bibitem[{\citenamefont{Mukhopadhyay et~al.}(2018)\citenamefont{Mukhopadhyay,
  Dehollain, Reichl, Wegscheider, and Vandersypen}}]{mukhopadhyay20182}
\bibinfo{author}{\bibfnamefont{U.}~\bibnamefont{Mukhopadhyay}},
  \bibinfo{author}{\bibfnamefont{J.~P.} \bibnamefont{Dehollain}},
  \bibinfo{author}{\bibfnamefont{C.}~\bibnamefont{Reichl}},
  \bibinfo{author}{\bibfnamefont{W.}~\bibnamefont{Wegscheider}},
  \bibnamefont{and} \bibinfo{author}{\bibfnamefont{L.~M.}
  \bibnamefont{Vandersypen}}, \bibinfo{journal}{Appl. Phys. Lett.}
  \textbf{\bibinfo{volume}{112}}, \bibinfo{pages}{183505}
  (\bibinfo{year}{2018}).

\bibitem[{\citenamefont{Hilker et~al.}(2017)\citenamefont{Hilker, Salomon,
  Grusdt, Omran, Boll, Demler, Bloch, and Gross}}]{hilker2017revealing}
\bibinfo{author}{\bibfnamefont{T.~A.} \bibnamefont{Hilker}},
  \bibinfo{author}{\bibfnamefont{G.}~\bibnamefont{Salomon}},
  \bibinfo{author}{\bibfnamefont{F.}~\bibnamefont{Grusdt}},
  \bibinfo{author}{\bibfnamefont{A.}~\bibnamefont{Omran}},
  \bibinfo{author}{\bibfnamefont{M.}~\bibnamefont{Boll}},
  \bibinfo{author}{\bibfnamefont{E.}~\bibnamefont{Demler}},
  \bibinfo{author}{\bibfnamefont{I.}~\bibnamefont{Bloch}}, \bibnamefont{and}
  \bibinfo{author}{\bibfnamefont{C.}~\bibnamefont{Gross}},
  \bibinfo{journal}{Science} \textbf{\bibinfo{volume}{357}},
  \bibinfo{pages}{484} (\bibinfo{year}{2017}).

\bibitem[{\citenamefont{Mazurenko et~al.}(2017)\citenamefont{Mazurenko, Chiu,
  Ji, Parsons, Kan{\'a}sz-Nagy, Schmidt, Grusdt, Demler, Greif, and
  Greiner}}]{mazurenko2017cold}
\bibinfo{author}{\bibfnamefont{A.}~\bibnamefont{Mazurenko}},
  \bibinfo{author}{\bibfnamefont{C.~S.} \bibnamefont{Chiu}},
  \bibinfo{author}{\bibfnamefont{G.}~\bibnamefont{Ji}},
  \bibinfo{author}{\bibfnamefont{M.~F.} \bibnamefont{Parsons}},
  \bibinfo{author}{\bibfnamefont{M.}~\bibnamefont{Kan{\'a}sz-Nagy}},
  \bibinfo{author}{\bibfnamefont{R.}~\bibnamefont{Schmidt}},
  \bibinfo{author}{\bibfnamefont{F.}~\bibnamefont{Grusdt}},
  \bibinfo{author}{\bibfnamefont{E.}~\bibnamefont{Demler}},
  \bibinfo{author}{\bibfnamefont{D.}~\bibnamefont{Greif}}, \bibnamefont{and}
  \bibinfo{author}{\bibfnamefont{M.}~\bibnamefont{Greiner}},
  \bibinfo{journal}{Nature} \textbf{\bibinfo{volume}{545}},
  \bibinfo{pages}{462} (\bibinfo{year}{2017}).

\bibitem[{\citenamefont{Barthelemy and
  Vandersypen}(2013)}]{barthelemy2013quantum}
\bibinfo{author}{\bibfnamefont{P.}~\bibnamefont{Barthelemy}} \bibnamefont{and}
  \bibinfo{author}{\bibfnamefont{L.~M.} \bibnamefont{Vandersypen}},
  \bibinfo{journal}{Ann. Phys.} \textbf{\bibinfo{volume}{525}},
  \bibinfo{pages}{808} (\bibinfo{year}{2013}).

\bibitem[{\citenamefont{Binkley et~al.}(1980)\citenamefont{Binkley, Pople, and
  Hehre}}]{binkley1980self}
\bibinfo{author}{\bibfnamefont{J.~S.} \bibnamefont{Binkley}},
  \bibinfo{author}{\bibfnamefont{J.~A.} \bibnamefont{Pople}}, \bibnamefont{and}
  \bibinfo{author}{\bibfnamefont{W.~J.} \bibnamefont{Hehre}},
  \bibinfo{journal}{J Am. Chem. Soc.} \textbf{\bibinfo{volume}{102}},
  \bibinfo{pages}{939} (\bibinfo{year}{1980}).

\bibitem[{\citenamefont{Ditchfield et~al.}(1971)\citenamefont{Ditchfield,
  Hehre, and Pople}}]{ditchfield1971self}
\bibinfo{author}{\bibfnamefont{R.}~\bibnamefont{Ditchfield}},
  \bibinfo{author}{\bibfnamefont{W.~J.} \bibnamefont{Hehre}}, \bibnamefont{and}
  \bibinfo{author}{\bibfnamefont{J.~A.} \bibnamefont{Pople}},
  \bibinfo{journal}{J Chem. Phys.} \textbf{\bibinfo{volume}{54}},
  \bibinfo{pages}{724} (\bibinfo{year}{1971}).

\bibitem[{\citenamefont{Frisch et~al.}(1984)\citenamefont{Frisch, Pople, and
  Binkley}}]{frisch1984self}
\bibinfo{author}{\bibfnamefont{M.~J.} \bibnamefont{Frisch}},
  \bibinfo{author}{\bibfnamefont{J.~A.} \bibnamefont{Pople}}, \bibnamefont{and}
  \bibinfo{author}{\bibfnamefont{J.~S.} \bibnamefont{Binkley}},
  \bibinfo{journal}{J Chem. Phys.} \textbf{\bibinfo{volume}{80}},
  \bibinfo{pages}{3265} (\bibinfo{year}{1984}).

\bibitem[{\citenamefont{Hariharan and Pople}(1973)}]{hariharan1973influence}
\bibinfo{author}{\bibfnamefont{P.~C.} \bibnamefont{Hariharan}}
  \bibnamefont{and} \bibinfo{author}{\bibfnamefont{J.~A.} \bibnamefont{Pople}},
  \bibinfo{journal}{Theor. Chim. Acta.} \textbf{\bibinfo{volume}{28}},
  \bibinfo{pages}{213} (\bibinfo{year}{1973}).

\bibitem[{\citenamefont{Hehre et~al.}(1969)\citenamefont{Hehre, Stewart, and
  Pople}}]{hehre1969self}
\bibinfo{author}{\bibfnamefont{W.~J.} \bibnamefont{Hehre}},
  \bibinfo{author}{\bibfnamefont{R.~F.} \bibnamefont{Stewart}},
  \bibnamefont{and} \bibinfo{author}{\bibfnamefont{J.~A.} \bibnamefont{Pople}},
  \bibinfo{journal}{J Chem. Phys.} \textbf{\bibinfo{volume}{51}},
  \bibinfo{pages}{2657} (\bibinfo{year}{1969}).

\bibitem[{\citenamefont{Hartree}(1928)}]{hartree1928wave}
\bibinfo{author}{\bibfnamefont{D.~R.} \bibnamefont{Hartree}},
  \bibinfo{journal}{Proc. Cambridge Philos. Soc.}
  \textbf{\bibinfo{volume}{24}}, \bibinfo{pages}{89} (\bibinfo{year}{1928}).

\bibitem[{\citenamefont{Hartree and Hartree}(1935)}]{hartree1935self}
\bibinfo{author}{\bibfnamefont{D.~R.} \bibnamefont{Hartree}} \bibnamefont{and}
  \bibinfo{author}{\bibfnamefont{W.}~\bibnamefont{Hartree}},
  \bibinfo{journal}{Proc. Roy. Soc. London Ser. A}
  \textbf{\bibinfo{volume}{150}}, \bibinfo{pages}{9} (\bibinfo{year}{1935}).

\bibitem[{\citenamefont{Fock}(1930{\natexlab{a}})}]{fock1930naherungsmethode}
\bibinfo{author}{\bibfnamefont{V.}~\bibnamefont{Fock}},
  \bibinfo{journal}{Zeitschrift f\"{u}r Physik} \textbf{\bibinfo{volume}{61}},
  \bibinfo{pages}{126} (\bibinfo{year}{1930}{\natexlab{a}}).

\bibitem[{\citenamefont{Fock}(1930{\natexlab{b}})}]{fock1930selfconsistent}
\bibinfo{author}{\bibfnamefont{V.}~\bibnamefont{Fock}},
  \bibinfo{journal}{Zeitschrift f\"{u}r Physik} \textbf{\bibinfo{volume}{62}},
  \bibinfo{pages}{795} (\bibinfo{year}{1930}{\natexlab{b}}).

\bibitem[{\citenamefont{Slater}(1930)}]{slater1930note}
\bibinfo{author}{\bibfnamefont{J.~C.} \bibnamefont{Slater}},
  \bibinfo{journal}{Phys. Rev.} \textbf{\bibinfo{volume}{35}},
  \bibinfo{pages}{210} (\bibinfo{year}{1930}).

\bibitem[{\citenamefont{Shavitt and Bartlett}(2009)}]{shavitt2009many}
\bibinfo{author}{\bibfnamefont{I.}~\bibnamefont{Shavitt}} \bibnamefont{and}
  \bibinfo{author}{\bibfnamefont{R.~J.} \bibnamefont{Bartlett}},
  \emph{\bibinfo{title}{Many-body methods in chemistry and physics: MBPT and
  coupled-cluster theory}} (\bibinfo{publisher}{Cambridge University Press,
  Cambridge, UK}, \bibinfo{year}{2009}).

\bibitem[{\citenamefont{Sherrill and
  Schaefer~III}(1999)}]{sherrill1999configuration}
\bibinfo{author}{\bibfnamefont{C.~D.} \bibnamefont{Sherrill}} \bibnamefont{and}
  \bibinfo{author}{\bibfnamefont{H.~F.} \bibnamefont{Schaefer~III}}, in
  \emph{\bibinfo{booktitle}{Adv. Quantum Chem.}}
  (\bibinfo{publisher}{Elsevier}, \bibinfo{year}{1999}),
  vol.~\bibinfo{volume}{34}, pp. \bibinfo{pages}{143--269}.

\bibitem[{\citenamefont{Pople and Head-Gordon}(1987)}]{pople1987CI}
\bibinfo{author}{\bibfnamefont{C.}~\bibnamefont{Pople}} \bibnamefont{and}
  \bibinfo{author}{\bibfnamefont{M.}~\bibnamefont{Head-Gordon}},
  \bibinfo{journal}{J. Chem. Phys.} \textbf{\bibinfo{volume}{87}},
  \bibinfo{pages}{5968} (\bibinfo{year}{1987}).

\bibitem[{\citenamefont{Hegarty and Robb}(1979)}]{hegarty1979application}
\bibinfo{author}{\bibfnamefont{D.}~\bibnamefont{Hegarty}} \bibnamefont{and}
  \bibinfo{author}{\bibfnamefont{M.~A.} \bibnamefont{Robb}},
  \bibinfo{journal}{Mol. Phys.} \textbf{\bibinfo{volume}{38}},
  \bibinfo{pages}{1795} (\bibinfo{year}{1979}).

\bibitem[{\citenamefont{Eade and Robb}(1981)}]{eade1981direct}
\bibinfo{author}{\bibfnamefont{R.~H.} \bibnamefont{Eade}} \bibnamefont{and}
  \bibinfo{author}{\bibfnamefont{M.~A.} \bibnamefont{Robb}},
  \bibinfo{journal}{Chem. Phys. Lett.} \textbf{\bibinfo{volume}{83}},
  \bibinfo{pages}{362} (\bibinfo{year}{1981}).

\bibitem[{\citenamefont{Yamamoto et~al.}(1996)\citenamefont{Yamamoto, Vreven,
  Robb, Frisch, and Schlegel}}]{yamamoto1996direct}
\bibinfo{author}{\bibfnamefont{N.}~\bibnamefont{Yamamoto}},
  \bibinfo{author}{\bibfnamefont{T.}~\bibnamefont{Vreven}},
  \bibinfo{author}{\bibfnamefont{M.~A.} \bibnamefont{Robb}},
  \bibinfo{author}{\bibfnamefont{M.~J.} \bibnamefont{Frisch}},
  \bibnamefont{and} \bibinfo{author}{\bibfnamefont{H.~B.}
  \bibnamefont{Schlegel}}, \bibinfo{journal}{Chem. Phys. Lett.}
  \textbf{\bibinfo{volume}{250}}, \bibinfo{pages}{373} (\bibinfo{year}{1996}).

\bibitem[{\citenamefont{Andersson et~al.}(1990)\citenamefont{Andersson,
  Malmqvist, Roos, Sadlej, and Wolinski}}]{andersson1990second}
\bibinfo{author}{\bibfnamefont{K.}~\bibnamefont{Andersson}},
  \bibinfo{author}{\bibfnamefont{P.~A.} \bibnamefont{Malmqvist}},
  \bibinfo{author}{\bibfnamefont{B.~O.} \bibnamefont{Roos}},
  \bibinfo{author}{\bibfnamefont{A.~J.} \bibnamefont{Sadlej}},
  \bibnamefont{and} \bibinfo{author}{\bibfnamefont{K.}~\bibnamefont{Wolinski}},
  \bibinfo{journal}{J. Phys. Chem.} \textbf{\bibinfo{volume}{94}},
  \bibinfo{pages}{5483} (\bibinfo{year}{1990}).

\bibitem[{\citenamefont{Andersson et~al.}(1992)\citenamefont{Andersson,
  Malmqvist, and Roos}}]{andersson1992second}
\bibinfo{author}{\bibfnamefont{K.}~\bibnamefont{Andersson}},
  \bibinfo{author}{\bibfnamefont{P.-{\AA}.} \bibnamefont{Malmqvist}},
  \bibnamefont{and} \bibinfo{author}{\bibfnamefont{B.~O.} \bibnamefont{Roos}},
  \bibinfo{journal}{J. Chem. Phys.} \textbf{\bibinfo{volume}{96}},
  \bibinfo{pages}{1218} (\bibinfo{year}{1992}).

\bibitem[{\citenamefont{Ufimtsev and
  Martinez}(2008{\natexlab{a}})}]{ufimtsev2008graphical}
\bibinfo{author}{\bibfnamefont{I.~S.} \bibnamefont{Ufimtsev}} \bibnamefont{and}
  \bibinfo{author}{\bibfnamefont{T.~J.} \bibnamefont{Martinez}},
  \bibinfo{journal}{Comput. Sci. Eng.} \textbf{\bibinfo{volume}{10}},
  \bibinfo{pages}{26} (\bibinfo{year}{2008}{\natexlab{a}}).

\bibitem[{\citenamefont{Ufimtsev and
  Martinez}(2008{\natexlab{b}})}]{ufimtsev2008quantum}
\bibinfo{author}{\bibfnamefont{I.~S.} \bibnamefont{Ufimtsev}} \bibnamefont{and}
  \bibinfo{author}{\bibfnamefont{T.~J.} \bibnamefont{Martinez}},
  \bibinfo{journal}{J Chem. Theory Comput.} \textbf{\bibinfo{volume}{4}},
  \bibinfo{pages}{222} (\bibinfo{year}{2008}{\natexlab{b}}).

\bibitem[{\citenamefont{DePrince~III and Hammond}(2011)}]{deprince2011coupled}
\bibinfo{author}{\bibfnamefont{A.~E.} \bibnamefont{DePrince~III}}
  \bibnamefont{and} \bibinfo{author}{\bibfnamefont{J.~R.}
  \bibnamefont{Hammond}}, \bibinfo{journal}{J. Chem. Theory Comput.}
  \textbf{\bibinfo{volume}{7}}, \bibinfo{pages}{1287} (\bibinfo{year}{2011}).

\bibitem[{\citenamefont{Hohenstein et~al.}(2015)\citenamefont{Hohenstein,
  Luehr, Ufimtsev, and Mart{\'\i}nez}}]{hohenstein2015atomic}
\bibinfo{author}{\bibfnamefont{E.~G.} \bibnamefont{Hohenstein}},
  \bibinfo{author}{\bibfnamefont{N.}~\bibnamefont{Luehr}},
  \bibinfo{author}{\bibfnamefont{I.~S.} \bibnamefont{Ufimtsev}},
  \bibnamefont{and} \bibinfo{author}{\bibfnamefont{T.~J.}
  \bibnamefont{Mart{\'\i}nez}}, \bibinfo{journal}{J. Chem. Phys.}
  \textbf{\bibinfo{volume}{142}}, \bibinfo{pages}{224103}
  (\bibinfo{year}{2015}).

\bibitem[{\citenamefont{Song and Mart{\'\i}nez}(2018)}]{song2018reduced}
\bibinfo{author}{\bibfnamefont{C.}~\bibnamefont{Song}} \bibnamefont{and}
  \bibinfo{author}{\bibfnamefont{T.~J.} \bibnamefont{Mart{\'\i}nez}},
  \bibinfo{journal}{J. Chem. Phys.} \textbf{\bibinfo{volume}{149}},
  \bibinfo{pages}{044108} (\bibinfo{year}{2018}).

\bibitem[{\citenamefont{Maksym and Chakraborty}(1990)}]{maksym1990quantum}
\bibinfo{author}{\bibfnamefont{P.}~\bibnamefont{Maksym}} \bibnamefont{and}
  \bibinfo{author}{\bibfnamefont{T.}~\bibnamefont{Chakraborty}},
  \bibinfo{journal}{Phys. Rev. Lett.} \textbf{\bibinfo{volume}{65}},
  \bibinfo{pages}{108} (\bibinfo{year}{1990}).

\bibitem[{\citenamefont{Merkt et~al.}(1991)\citenamefont{Merkt, Huser, and
  Wagner}}]{merkt1991energy}
\bibinfo{author}{\bibfnamefont{U.}~\bibnamefont{Merkt}},
  \bibinfo{author}{\bibfnamefont{J.}~\bibnamefont{Huser}}, \bibnamefont{and}
  \bibinfo{author}{\bibfnamefont{M.}~\bibnamefont{Wagner}},
  \bibinfo{journal}{Phys. Rev. B} \textbf{\bibinfo{volume}{43}},
  \bibinfo{pages}{7320} (\bibinfo{year}{1991}).

\bibitem[{\citenamefont{Pfannkuche et~al.}(1993)\citenamefont{Pfannkuche,
  Gudmundsson, and Maksym}}]{pfannkuche1993comparison}
\bibinfo{author}{\bibfnamefont{D.}~\bibnamefont{Pfannkuche}},
  \bibinfo{author}{\bibfnamefont{V.}~\bibnamefont{Gudmundsson}},
  \bibnamefont{and} \bibinfo{author}{\bibfnamefont{P.~A.}
  \bibnamefont{Maksym}}, \bibinfo{journal}{Phys. Rev. B}
  \textbf{\bibinfo{volume}{47}}, \bibinfo{pages}{2244} (\bibinfo{year}{1993}).

\bibitem[{\citenamefont{Hawrylak and
  Pfannkuche}(1993)}]{hawrylak1993magnetoluminescence}
\bibinfo{author}{\bibfnamefont{P.}~\bibnamefont{Hawrylak}} \bibnamefont{and}
  \bibinfo{author}{\bibfnamefont{D.}~\bibnamefont{Pfannkuche}},
  \bibinfo{journal}{Phys. Rev. Lett.} \textbf{\bibinfo{volume}{70}},
  \bibinfo{pages}{485} (\bibinfo{year}{1993}).

\bibitem[{\citenamefont{Yang et~al.}(1993)\citenamefont{Yang, MacDonald, and
  Johnson}}]{yang1993addition}
\bibinfo{author}{\bibfnamefont{S.-R.~E.} \bibnamefont{Yang}},
  \bibinfo{author}{\bibfnamefont{A.}~\bibnamefont{MacDonald}},
  \bibnamefont{and} \bibinfo{author}{\bibfnamefont{M.}~\bibnamefont{Johnson}},
  \bibinfo{journal}{Phys. Rev. Lett.} \textbf{\bibinfo{volume}{71}},
  \bibinfo{pages}{3194} (\bibinfo{year}{1993}).

\bibitem[{\citenamefont{Hawrylak}(1993)}]{hawrylak1993single}
\bibinfo{author}{\bibfnamefont{P.}~\bibnamefont{Hawrylak}},
  \bibinfo{journal}{Phys. Rev. Lett.} \textbf{\bibinfo{volume}{71}},
  \bibinfo{pages}{3347} (\bibinfo{year}{1993}).

\bibitem[{\citenamefont{Palacios et~al.}(1994)\citenamefont{Palacios,
  Martin-Moreno, Chiappe, Louis, and Tejedor}}]{palacios1994capacitance}
\bibinfo{author}{\bibfnamefont{J.}~\bibnamefont{Palacios}},
  \bibinfo{author}{\bibfnamefont{L.}~\bibnamefont{Martin-Moreno}},
  \bibinfo{author}{\bibfnamefont{G.}~\bibnamefont{Chiappe}},
  \bibinfo{author}{\bibfnamefont{E.}~\bibnamefont{Louis}}, \bibnamefont{and}
  \bibinfo{author}{\bibfnamefont{C.}~\bibnamefont{Tejedor}},
  \bibinfo{journal}{Phys. Rev. B} \textbf{\bibinfo{volume}{50}},
  \bibinfo{pages}{5760} (\bibinfo{year}{1994}).

\bibitem[{\citenamefont{Wojs and Hawrylak}(1995)}]{wojs1995negatively}
\bibinfo{author}{\bibfnamefont{A.}~\bibnamefont{Wojs}} \bibnamefont{and}
  \bibinfo{author}{\bibfnamefont{P.}~\bibnamefont{Hawrylak}},
  \bibinfo{journal}{Phys. Rev. B} \textbf{\bibinfo{volume}{51}},
  \bibinfo{pages}{10880} (\bibinfo{year}{1995}).

\bibitem[{\citenamefont{Oaknin et~al.}(1995)\citenamefont{Oaknin,
  Martin-Moreno, Palacios, and Tejedor}}]{oaknin1995low}
\bibinfo{author}{\bibfnamefont{J.}~\bibnamefont{Oaknin}},
  \bibinfo{author}{\bibfnamefont{L.}~\bibnamefont{Martin-Moreno}},
  \bibinfo{author}{\bibfnamefont{J.}~\bibnamefont{Palacios}}, \bibnamefont{and}
  \bibinfo{author}{\bibfnamefont{C.}~\bibnamefont{Tejedor}},
  \bibinfo{journal}{Phys. Rev. Lett.} \textbf{\bibinfo{volume}{74}},
  \bibinfo{pages}{5120} (\bibinfo{year}{1995}).

\bibitem[{\citenamefont{Wojs and Hawrylak}(1996)}]{wojs1996charging}
\bibinfo{author}{\bibfnamefont{A.}~\bibnamefont{Wojs}} \bibnamefont{and}
  \bibinfo{author}{\bibfnamefont{P.}~\bibnamefont{Hawrylak}},
  \bibinfo{journal}{Phys. Rev. B} \textbf{\bibinfo{volume}{53}},
  \bibinfo{pages}{10841} (\bibinfo{year}{1996}).

\bibitem[{\citenamefont{Maksym}(1996)}]{maksym1996eckardt}
\bibinfo{author}{\bibfnamefont{P.}~\bibnamefont{Maksym}},
  \bibinfo{journal}{Phys. Rev. B} \textbf{\bibinfo{volume}{53}},
  \bibinfo{pages}{10871} (\bibinfo{year}{1996}).

\bibitem[{\citenamefont{Hawrylak et~al.}(1996)\citenamefont{Hawrylak, Wojs, and
  Brum}}]{hawrylak1996magnetoexcitons}
\bibinfo{author}{\bibfnamefont{P.}~\bibnamefont{Hawrylak}},
  \bibinfo{author}{\bibfnamefont{A.}~\bibnamefont{Wojs}}, \bibnamefont{and}
  \bibinfo{author}{\bibfnamefont{J.~A.} \bibnamefont{Brum}},
  \bibinfo{journal}{Phys. Rev. B} \textbf{\bibinfo{volume}{54}},
  \bibinfo{pages}{11397} (\bibinfo{year}{1996}).

\bibitem[{\citenamefont{Wojs and
  Hawrylak}(1997{\natexlab{a}})}]{wojs1997theory}
\bibinfo{author}{\bibfnamefont{A.}~\bibnamefont{Wojs}} \bibnamefont{and}
  \bibinfo{author}{\bibfnamefont{P.}~\bibnamefont{Hawrylak}},
  \bibinfo{journal}{Phys. Rev. B} \textbf{\bibinfo{volume}{55}},
  \bibinfo{pages}{13066} (\bibinfo{year}{1997}{\natexlab{a}}).

\bibitem[{\citenamefont{Wojs and
  Hawrylak}(1997{\natexlab{b}})}]{wojs1997spectral}
\bibinfo{author}{\bibfnamefont{A.}~\bibnamefont{Wojs}} \bibnamefont{and}
  \bibinfo{author}{\bibfnamefont{P.}~\bibnamefont{Hawrylak}},
  \bibinfo{journal}{Phys. Rev. B} \textbf{\bibinfo{volume}{56}},
  \bibinfo{pages}{13227} (\bibinfo{year}{1997}{\natexlab{b}}).

\bibitem[{\citenamefont{Eto}(1997)}]{eto1997electronic}
\bibinfo{author}{\bibfnamefont{M.}~\bibnamefont{Eto}}, \bibinfo{journal}{Jpn.
  J. Appl. Phys.} \textbf{\bibinfo{volume}{36}}, \bibinfo{pages}{3924}
  (\bibinfo{year}{1997}).

\bibitem[{\citenamefont{Maksym}(1998)}]{maksym1998quantum}
\bibinfo{author}{\bibfnamefont{P.}~\bibnamefont{Maksym}},
  \bibinfo{journal}{Physica B} \textbf{\bibinfo{volume}{249}},
  \bibinfo{pages}{233} (\bibinfo{year}{1998}).

\bibitem[{\citenamefont{Reimann et~al.}(2000)\citenamefont{Reimann, Koskinen,
  and Manninen}}]{reimann2000formation}
\bibinfo{author}{\bibfnamefont{S.}~\bibnamefont{Reimann}},
  \bibinfo{author}{\bibfnamefont{M.}~\bibnamefont{Koskinen}}, \bibnamefont{and}
  \bibinfo{author}{\bibfnamefont{M.}~\bibnamefont{Manninen}},
  \bibinfo{journal}{Phys. Rev. B} \textbf{\bibinfo{volume}{62}},
  \bibinfo{pages}{8108} (\bibinfo{year}{2000}).

\bibitem[{\citenamefont{Mikhailov}(2002)}]{mikhailov2002quantum}
\bibinfo{author}{\bibfnamefont{S.~A.} \bibnamefont{Mikhailov}},
  \bibinfo{journal}{Phys. Rev. B} \textbf{\bibinfo{volume}{65}},
  \bibinfo{pages}{115312} (\bibinfo{year}{2002}).

\bibitem[{\citenamefont{Bolton}(1996)}]{bolton1996fixed}
\bibinfo{author}{\bibfnamefont{F.}~\bibnamefont{Bolton}},
  \bibinfo{journal}{Phys. Rev. B} \textbf{\bibinfo{volume}{54}},
  \bibinfo{pages}{4780} (\bibinfo{year}{1996}).

\bibitem[{\citenamefont{Harju et~al.}(1999{\natexlab{a}})\citenamefont{Harju,
  Sverdlov, Nieminen, and Halonen}}]{harju1999many}
\bibinfo{author}{\bibfnamefont{A.}~\bibnamefont{Harju}},
  \bibinfo{author}{\bibfnamefont{V.}~\bibnamefont{Sverdlov}},
  \bibinfo{author}{\bibfnamefont{R.~M.} \bibnamefont{Nieminen}},
  \bibnamefont{and} \bibinfo{author}{\bibfnamefont{V.}~\bibnamefont{Halonen}},
  \bibinfo{journal}{Phys. Rev. B} \textbf{\bibinfo{volume}{59}},
  \bibinfo{pages}{5622} (\bibinfo{year}{1999}{\natexlab{a}}).

\bibitem[{\citenamefont{Harju et~al.}(1999{\natexlab{b}})\citenamefont{Harju,
  Siljam{\"a}ki, and Nieminen}}]{harju1999wave}
\bibinfo{author}{\bibfnamefont{A.}~\bibnamefont{Harju}},
  \bibinfo{author}{\bibfnamefont{S.}~\bibnamefont{Siljam{\"a}ki}},
  \bibnamefont{and} \bibinfo{author}{\bibfnamefont{R.~M.}
  \bibnamefont{Nieminen}}, \bibinfo{journal}{Phys. Rev. B}
  \textbf{\bibinfo{volume}{60}}, \bibinfo{pages}{1807}
  (\bibinfo{year}{1999}{\natexlab{b}}).

\bibitem[{\citenamefont{Harju et~al.}(2002)\citenamefont{Harju, Siljam{\"a}ki,
  and Nieminen}}]{harju2002wigner}
\bibinfo{author}{\bibfnamefont{A.}~\bibnamefont{Harju}},
  \bibinfo{author}{\bibfnamefont{S.}~\bibnamefont{Siljam{\"a}ki}},
  \bibnamefont{and} \bibinfo{author}{\bibfnamefont{R.~M.}
  \bibnamefont{Nieminen}}, \bibinfo{journal}{Phys. Rev. B}
  \textbf{\bibinfo{volume}{65}}, \bibinfo{pages}{075309}
  (\bibinfo{year}{2002}).

\bibitem[{\citenamefont{Siljam\"aki et~al.}(2002)\citenamefont{Siljam\"aki,
  Harju, Nieminen, Sverdlov, and Hyv\"onen}}]{siljamaki2002various}
\bibinfo{author}{\bibfnamefont{S.}~\bibnamefont{Siljam\"aki}},
  \bibinfo{author}{\bibfnamefont{A.}~\bibnamefont{Harju}},
  \bibinfo{author}{\bibfnamefont{R.~M.} \bibnamefont{Nieminen}},
  \bibinfo{author}{\bibfnamefont{V.~A.} \bibnamefont{Sverdlov}},
  \bibnamefont{and}
  \bibinfo{author}{\bibfnamefont{P.}~\bibnamefont{Hyv\"onen}},
  \bibinfo{journal}{Phys. Rev. B} \textbf{\bibinfo{volume}{65}},
  \bibinfo{pages}{121306} (\bibinfo{year}{2002}).

\bibitem[{\citenamefont{Burkard et~al.}(1999)\citenamefont{Burkard, Loss, and
  DiVincenzo}}]{burkard1999coupled}
\bibinfo{author}{\bibfnamefont{G.}~\bibnamefont{Burkard}},
  \bibinfo{author}{\bibfnamefont{D.}~\bibnamefont{Loss}}, \bibnamefont{and}
  \bibinfo{author}{\bibfnamefont{D.~P.} \bibnamefont{DiVincenzo}},
  \bibinfo{journal}{Phys. Rev. B} \textbf{\bibinfo{volume}{59}},
  \bibinfo{pages}{2070} (\bibinfo{year}{1999}).

\bibitem[{\citenamefont{Li et~al.}(2009)\citenamefont{Li, Yannouleas, and
  Landman}}]{li2009artificial}
\bibinfo{author}{\bibfnamefont{Y.}~\bibnamefont{Li}},
  \bibinfo{author}{\bibfnamefont{C.}~\bibnamefont{Yannouleas}},
  \bibnamefont{and} \bibinfo{author}{\bibfnamefont{U.}~\bibnamefont{Landman}},
  \bibinfo{journal}{Phys. Rev. B} \textbf{\bibinfo{volume}{80}},
  \bibinfo{pages}{045326} (\bibinfo{year}{2009}).

\bibitem[{\citenamefont{Nielsen et~al.}(2012)\citenamefont{Nielsen, Rahman, and
  Muller}}]{nielsen2012many}
\bibinfo{author}{\bibfnamefont{E.}~\bibnamefont{Nielsen}},
  \bibinfo{author}{\bibfnamefont{R.}~\bibnamefont{Rahman}}, \bibnamefont{and}
  \bibinfo{author}{\bibfnamefont{R.~P.} \bibnamefont{Muller}},
  \bibinfo{journal}{J. Appl. Phys.} \textbf{\bibinfo{volume}{112}},
  \bibinfo{pages}{114304} (\bibinfo{year}{2012}).

\bibitem[{\citenamefont{Stepanenko et~al.}(2012)\citenamefont{Stepanenko,
  Rudner, Halperin, and Loss}}]{stepanenko2012singlet}
\bibinfo{author}{\bibfnamefont{D.}~\bibnamefont{Stepanenko}},
  \bibinfo{author}{\bibfnamefont{M.}~\bibnamefont{Rudner}},
  \bibinfo{author}{\bibfnamefont{B.~I.} \bibnamefont{Halperin}},
  \bibnamefont{and} \bibinfo{author}{\bibfnamefont{D.}~\bibnamefont{Loss}},
  \bibinfo{journal}{Phys. Rev. B} \textbf{\bibinfo{volume}{85}},
  \bibinfo{pages}{075416} (\bibinfo{year}{2012}).

\bibitem[{\citenamefont{Yannouleas et~al.}(2016)\citenamefont{Yannouleas,
  Brandt, and Landman}}]{yannouleas2016ultracold}
\bibinfo{author}{\bibfnamefont{C.}~\bibnamefont{Yannouleas}},
  \bibinfo{author}{\bibfnamefont{B.~B.} \bibnamefont{Brandt}},
  \bibnamefont{and} \bibinfo{author}{\bibfnamefont{U.}~\bibnamefont{Landman}},
  \bibinfo{journal}{New J. Phys.} \textbf{\bibinfo{volume}{18}},
  \bibinfo{pages}{073018} (\bibinfo{year}{2016}).

\bibitem[{\citenamefont{Brandt et~al.}(2017)\citenamefont{Brandt, Yannouleas,
  and Landman}}]{brandt2017bottom}
\bibinfo{author}{\bibfnamefont{B.~B.} \bibnamefont{Brandt}},
  \bibinfo{author}{\bibfnamefont{C.}~\bibnamefont{Yannouleas}},
  \bibnamefont{and} \bibinfo{author}{\bibfnamefont{U.}~\bibnamefont{Landman}},
  \bibinfo{journal}{Phys. Rev. A} \textbf{\bibinfo{volume}{95}},
  \bibinfo{pages}{043617} (\bibinfo{year}{2017}).

\bibitem[{\citenamefont{Abolfath and Hawrylak}(2006)}]{abolfath2006real}
\bibinfo{author}{\bibfnamefont{R.~M.} \bibnamefont{Abolfath}} \bibnamefont{and}
  \bibinfo{author}{\bibfnamefont{P.}~\bibnamefont{Hawrylak}},
  \bibinfo{journal}{J. Chem. Phys.} \textbf{\bibinfo{volume}{125}},
  \bibinfo{pages}{034707} (\bibinfo{year}{2006}).

\bibitem[{\citenamefont{Stopa et~al.}(2006)\citenamefont{Stopa, Vidan, Hatano,
  Tarucha, and Westervelt}}]{stopa2006electronic}
\bibinfo{author}{\bibfnamefont{M.}~\bibnamefont{Stopa}},
  \bibinfo{author}{\bibfnamefont{A.}~\bibnamefont{Vidan}},
  \bibinfo{author}{\bibfnamefont{T.}~\bibnamefont{Hatano}},
  \bibinfo{author}{\bibfnamefont{S.}~\bibnamefont{Tarucha}}, \bibnamefont{and}
  \bibinfo{author}{\bibfnamefont{R.}~\bibnamefont{Westervelt}},
  \bibinfo{journal}{Physica E} \textbf{\bibinfo{volume}{34}},
  \bibinfo{pages}{616} (\bibinfo{year}{2006}).

\bibitem[{\citenamefont{Abolfath et~al.}(2006)\citenamefont{Abolfath, Dybalski,
  and Hawrylak}}]{abolfath2006theory}
\bibinfo{author}{\bibfnamefont{R.~M.} \bibnamefont{Abolfath}},
  \bibinfo{author}{\bibfnamefont{W.}~\bibnamefont{Dybalski}}, \bibnamefont{and}
  \bibinfo{author}{\bibfnamefont{P.}~\bibnamefont{Hawrylak}},
  \bibinfo{journal}{Phys. Rev. B} \textbf{\bibinfo{volume}{73}},
  \bibinfo{pages}{075314} (\bibinfo{year}{2006}).

\bibitem[{\citenamefont{Stafford and Sarma}(1994)}]{stafford1994collective}
\bibinfo{author}{\bibfnamefont{C.~A.} \bibnamefont{Stafford}} \bibnamefont{and}
  \bibinfo{author}{\bibfnamefont{S.~D.} \bibnamefont{Sarma}},
  \bibinfo{journal}{Phys. Rev. Lett.} \textbf{\bibinfo{volume}{72}},
  \bibinfo{pages}{3590} (\bibinfo{year}{1994}).

\bibitem[{\citenamefont{Stafford and Sarma}(1997)}]{stafford1997coherent}
\bibinfo{author}{\bibfnamefont{C.~A.} \bibnamefont{Stafford}} \bibnamefont{and}
  \bibinfo{author}{\bibfnamefont{S.~D.} \bibnamefont{Sarma}},
  \bibinfo{journal}{Phys. Lett. A} \textbf{\bibinfo{volume}{230}},
  \bibinfo{pages}{73} (\bibinfo{year}{1997}).

\bibitem[{\citenamefont{Stafford et~al.}(1998)\citenamefont{Stafford, Kotlyar,
  and Sarma}}]{stafford1998coherent}
\bibinfo{author}{\bibfnamefont{C.~A.} \bibnamefont{Stafford}},
  \bibinfo{author}{\bibfnamefont{R.}~\bibnamefont{Kotlyar}}, \bibnamefont{and}
  \bibinfo{author}{\bibfnamefont{S.~D.} \bibnamefont{Sarma}},
  \bibinfo{journal}{Phys. Rev. B} \textbf{\bibinfo{volume}{58}},
  \bibinfo{pages}{7091} (\bibinfo{year}{1998}).

\bibitem[{\citenamefont{Kotlyar
  et~al.}(1998{\natexlab{a}})\citenamefont{Kotlyar, Stafford, and
  Sarma}}]{kotlyar1998addition}
\bibinfo{author}{\bibfnamefont{R.}~\bibnamefont{Kotlyar}},
  \bibinfo{author}{\bibfnamefont{C.}~\bibnamefont{Stafford}}, \bibnamefont{and}
  \bibinfo{author}{\bibfnamefont{S.~D.} \bibnamefont{Sarma}},
  \bibinfo{journal}{Phys. Rev. B} \textbf{\bibinfo{volume}{58}},
  \bibinfo{pages}{3989} (\bibinfo{year}{1998}{\natexlab{a}}).

\bibitem[{\citenamefont{Kotlyar
  et~al.}(1998{\natexlab{b}})\citenamefont{Kotlyar, Stafford, and
  Sarma}}]{kotlyar1998correlated}
\bibinfo{author}{\bibfnamefont{R.}~\bibnamefont{Kotlyar}},
  \bibinfo{author}{\bibfnamefont{C.}~\bibnamefont{Stafford}}, \bibnamefont{and}
  \bibinfo{author}{\bibfnamefont{S.~D.} \bibnamefont{Sarma}},
  \bibinfo{journal}{Phys. Rev. B} \textbf{\bibinfo{volume}{58}},
  \bibinfo{pages}{R1746} (\bibinfo{year}{1998}{\natexlab{b}}).

\bibitem[{\citenamefont{Dehollain et~al.}(2019)\citenamefont{Dehollain,
  Mukhopadhyay, Michal, Wang, Wunsch, Reichl, Wegscheider, Rudner, Demler, and
  Vandersypen}}]{dehollain2019nagaoka}
\bibinfo{author}{\bibfnamefont{J.~P.} \bibnamefont{Dehollain}},
  \bibinfo{author}{\bibfnamefont{U.}~\bibnamefont{Mukhopadhyay}},
  \bibinfo{author}{\bibfnamefont{V.~P.} \bibnamefont{Michal}},
  \bibinfo{author}{\bibfnamefont{Y.}~\bibnamefont{Wang}},
  \bibinfo{author}{\bibfnamefont{B.}~\bibnamefont{Wunsch}},
  \bibinfo{author}{\bibfnamefont{C.}~\bibnamefont{Reichl}},
  \bibinfo{author}{\bibfnamefont{W.}~\bibnamefont{Wegscheider}},
  \bibinfo{author}{\bibfnamefont{M.~S.} \bibnamefont{Rudner}},
  \bibinfo{author}{\bibfnamefont{E.}~\bibnamefont{Demler}}, \bibnamefont{and}
  \bibinfo{author}{\bibfnamefont{L.~M.~K.} \bibnamefont{Vandersypen}},
  \bibinfo{journal}{arXiv:1904.05680}  (\bibinfo{year}{2019}).

\bibitem[{\citenamefont{Nagaoka}(1966)}]{nagaoka1966ferromagnetism}
\bibinfo{author}{\bibfnamefont{Y.}~\bibnamefont{Nagaoka}},
  \bibinfo{journal}{Phys. Rev.} \textbf{\bibinfo{volume}{147}},
  \bibinfo{pages}{392} (\bibinfo{year}{1966}).

\bibitem[{pot()}]{potentialshape}
\bibinfo{note}{A more realistic treatment of the potential landscape lies in
  solving the Poisson equation given by the geometry of the electrodes.}

\bibitem[{\citenamefont{van Beveren et~al.}(2005)\citenamefont{van Beveren,
  Hanson, Vink, Koppens, Kouwenhoven, and Vandersypen}}]{van2005spin}
\bibinfo{author}{\bibfnamefont{L.~W.} \bibnamefont{van Beveren}},
  \bibinfo{author}{\bibfnamefont{R.}~\bibnamefont{Hanson}},
  \bibinfo{author}{\bibfnamefont{I.}~\bibnamefont{Vink}},
  \bibinfo{author}{\bibfnamefont{F.}~\bibnamefont{Koppens}},
  \bibinfo{author}{\bibfnamefont{L.}~\bibnamefont{Kouwenhoven}},
  \bibnamefont{and}
  \bibinfo{author}{\bibfnamefont{L.}~\bibnamefont{Vandersypen}},
  \bibinfo{journal}{New J. Phys.} \textbf{\bibinfo{volume}{7}},
  \bibinfo{pages}{182} (\bibinfo{year}{2005}).

\bibitem[{\citenamefont{Szabo and Ostlund}(2012)}]{szabo2012modern}
\bibinfo{author}{\bibfnamefont{A.}~\bibnamefont{Szabo}} \bibnamefont{and}
  \bibinfo{author}{\bibfnamefont{N.~S.} \bibnamefont{Ostlund}},
  \emph{\bibinfo{title}{Modern quantum chemistry: introduction to advanced
  electronic structure theory}} (\bibinfo{publisher}{Dover Publications,
  Mineola, New York}, \bibinfo{year}{2012}).

\bibitem[{\citenamefont{Helgaker et~al.}(2014)\citenamefont{Helgaker,
  Jorgensen, and Olsen}}]{helgaker2014molecular}
\bibinfo{author}{\bibfnamefont{T.}~\bibnamefont{Helgaker}},
  \bibinfo{author}{\bibfnamefont{P.}~\bibnamefont{Jorgensen}},
  \bibnamefont{and} \bibinfo{author}{\bibfnamefont{J.}~\bibnamefont{Olsen}},
  \emph{\bibinfo{title}{Molecular electronic-structure theory}}
  (\bibinfo{publisher}{John Wiley \& Sons, New York}, \bibinfo{year}{2014}).

\bibitem[{res()}]{resonance}
\bibinfo{note}{Here we refer to a process which does not change the total site
  energy as a resonant process.}

\bibitem[{non()}]{nonresonance}
\bibinfo{note}{Note, these terms become resonant when using the parabolic
  potential and should be considered in the
  calculation\cite{sheng2005multiband}. Even for a finite potential well, with
  the smaller level spacing these terms may still be important through virtual
  process. We will analyze their effect in a separate publication.}

\bibitem[{\citenamefont{Dagotto}(2013)}]{dagotto2013nanoscale}
\bibinfo{author}{\bibfnamefont{E.}~\bibnamefont{Dagotto}},
  \emph{\bibinfo{title}{Nanoscale phase separation and colossal
  magnetoresistance: the physics of manganites and related compounds}}, vol.
  \bibinfo{volume}{136} (\bibinfo{publisher}{Springer Science \& Business
  Media, Berlin}, \bibinfo{year}{2013}).

\bibitem[{\citenamefont{von Stecher et~al.}(2010)\citenamefont{von Stecher,
  Demler, Lukin, and Rey}}]{von2010probing}
\bibinfo{author}{\bibfnamefont{J.}~\bibnamefont{von Stecher}},
  \bibinfo{author}{\bibfnamefont{E.}~\bibnamefont{Demler}},
  \bibinfo{author}{\bibfnamefont{M.~D.} \bibnamefont{Lukin}}, \bibnamefont{and}
  \bibinfo{author}{\bibfnamefont{A.~M.} \bibnamefont{Rey}},
  \bibinfo{journal}{New J. Phys.} \textbf{\bibinfo{volume}{12}},
  \bibinfo{pages}{055009} (\bibinfo{year}{2010}).

\bibitem[{coh()}]{coherence}
\bibinfo{note}{To avoid possible confusion, we emphasize that these local
  orbitals can be physical in a system that is assembled dynamically from
  single dots. In the experiment we are modeling, only the global orbital --
  the eigenstate of the many-body Hamiltonian
  Eq.~\eqref{eq:nonTighBindingHamiltonian} -- is intrinsic to the system.
  Therefore, the coherence of these conceptual hopping paths is not relevant.}

\bibitem[{con()}]{convergence}
\bibinfo{note}{A benchmark with a 14-orbital truncation indicates the relative
  error being less than 1\%.}

\bibitem[{\citenamefont{Lehoucq et~al.}(1998)\citenamefont{Lehoucq, Sorensen,
  and Yang}}]{lehoucq1998arpack}
\bibinfo{author}{\bibfnamefont{R.~B.} \bibnamefont{Lehoucq}},
  \bibinfo{author}{\bibfnamefont{D.~C.} \bibnamefont{Sorensen}},
  \bibnamefont{and} \bibinfo{author}{\bibfnamefont{C.}~\bibnamefont{Yang}},
  \emph{\bibinfo{title}{ARPACK Users' Guide: Solution of Large-Scale Eigenvalue
  Problems with Implicitly Restarted Arnoldi Methods}}
  (\bibinfo{publisher}{SIAM, Philadelphia}, \bibinfo{year}{1998}).

\bibitem[{\citenamefont{Jia et~al.}(2018)\citenamefont{Jia, Wang, Mendl,
  Moritz, and Devereaux}}]{jia2017paradeisos}
\bibinfo{author}{\bibfnamefont{C.}~\bibnamefont{Jia}},
  \bibinfo{author}{\bibfnamefont{Y.}~\bibnamefont{Wang}},
  \bibinfo{author}{\bibfnamefont{C.}~\bibnamefont{Mendl}},
  \bibinfo{author}{\bibfnamefont{B.}~\bibnamefont{Moritz}}, \bibnamefont{and}
  \bibinfo{author}{\bibfnamefont{T.}~\bibnamefont{Devereaux}},
  \bibinfo{journal}{Comput. Phys. Commun.} \textbf{\bibinfo{volume}{224}},
  \bibinfo{pages}{81} (\bibinfo{year}{2018}).

\bibitem[{\citenamefont{Mattis}(2003)}]{mattis2003eigenvalues}
\bibinfo{author}{\bibfnamefont{D.}~\bibnamefont{Mattis}},
  \bibinfo{journal}{Int. J. Nanosci.} \textbf{\bibinfo{volume}{2}},
  \bibinfo{pages}{165} (\bibinfo{year}{2003}).

\bibitem[{\citenamefont{Kollar et~al.}(1996)\citenamefont{Kollar, Strack, and
  Vollhardt}}]{kollar1996ferromagnetism}
\bibinfo{author}{\bibfnamefont{M.}~\bibnamefont{Kollar}},
  \bibinfo{author}{\bibfnamefont{R.}~\bibnamefont{Strack}}, \bibnamefont{and}
  \bibinfo{author}{\bibfnamefont{D.}~\bibnamefont{Vollhardt}},
  \bibinfo{journal}{Phys. Rev. B} \textbf{\bibinfo{volume}{53}},
  \bibinfo{pages}{9225} (\bibinfo{year}{1996}).

\bibitem[{\citenamefont{Petta et~al.}(2005)\citenamefont{Petta, Johnson,
  Taylor, Laird, Yacoby, Lukin, Marcus, Hanson, and
  Gossard}}]{petta2005coherent}
\bibinfo{author}{\bibfnamefont{J.~R.} \bibnamefont{Petta}},
  \bibinfo{author}{\bibfnamefont{A.~C.} \bibnamefont{Johnson}},
  \bibinfo{author}{\bibfnamefont{J.~M.} \bibnamefont{Taylor}},
  \bibinfo{author}{\bibfnamefont{E.~A.} \bibnamefont{Laird}},
  \bibinfo{author}{\bibfnamefont{A.}~\bibnamefont{Yacoby}},
  \bibinfo{author}{\bibfnamefont{M.~D.} \bibnamefont{Lukin}},
  \bibinfo{author}{\bibfnamefont{C.~M.} \bibnamefont{Marcus}},
  \bibinfo{author}{\bibfnamefont{M.~P.} \bibnamefont{Hanson}},
  \bibnamefont{and} \bibinfo{author}{\bibfnamefont{A.~C.}
  \bibnamefont{Gossard}}, \bibinfo{journal}{Science}
  \textbf{\bibinfo{volume}{309}}, \bibinfo{pages}{2180} (\bibinfo{year}{2005}).

\bibitem[{\citenamefont{Lieb}(1962)}]{lieb1962lieb}
\bibinfo{author}{\bibfnamefont{E.}~\bibnamefont{Lieb}}, \bibinfo{journal}{Phys.
  Rev.} \textbf{\bibinfo{volume}{125}}, \bibinfo{pages}{164}
  (\bibinfo{year}{1962}).

\bibitem[{\citenamefont{Nelson et~al.}(2004)\citenamefont{Nelson, Mao, Maeno,
  and Liu}}]{nelson2004odd}
\bibinfo{author}{\bibfnamefont{K.}~\bibnamefont{Nelson}},
  \bibinfo{author}{\bibfnamefont{Z.}~\bibnamefont{Mao}},
  \bibinfo{author}{\bibfnamefont{Y.}~\bibnamefont{Maeno}}, \bibnamefont{and}
  \bibinfo{author}{\bibfnamefont{Y.}~\bibnamefont{Liu}},
  \bibinfo{journal}{Science} \textbf{\bibinfo{volume}{306}},
  \bibinfo{pages}{1151} (\bibinfo{year}{2004}).

\bibitem[{\citenamefont{Wu}(2008)}]{wu2008orbital}
\bibinfo{author}{\bibfnamefont{C.}~\bibnamefont{Wu}}, \bibinfo{journal}{Phys.
  Rev. Lett.} \textbf{\bibinfo{volume}{100}}, \bibinfo{pages}{200406}
  (\bibinfo{year}{2008}).

\bibitem[{\citenamefont{Kim et~al.}(2009)\citenamefont{Kim, Ohsumi, Komesu,
  Sakai, Morita, Takagi, and Arima}}]{kim2009phase}
\bibinfo{author}{\bibfnamefont{B.}~\bibnamefont{Kim}},
  \bibinfo{author}{\bibfnamefont{H.}~\bibnamefont{Ohsumi}},
  \bibinfo{author}{\bibfnamefont{T.}~\bibnamefont{Komesu}},
  \bibinfo{author}{\bibfnamefont{S.}~\bibnamefont{Sakai}},
  \bibinfo{author}{\bibfnamefont{T.}~\bibnamefont{Morita}},
  \bibinfo{author}{\bibfnamefont{H.}~\bibnamefont{Takagi}}, \bibnamefont{and}
  \bibinfo{author}{\bibfnamefont{T.-h.} \bibnamefont{Arima}},
  \bibinfo{journal}{Science} \textbf{\bibinfo{volume}{323}},
  \bibinfo{pages}{1329} (\bibinfo{year}{2009}).

\bibitem[{\citenamefont{Sheng et~al.}(2005)\citenamefont{Sheng, Cheng, and
  Hawrylak}}]{sheng2005multiband}
\bibinfo{author}{\bibfnamefont{W.}~\bibnamefont{Sheng}},
  \bibinfo{author}{\bibfnamefont{S.-J.} \bibnamefont{Cheng}}, \bibnamefont{and}
  \bibinfo{author}{\bibfnamefont{P.}~\bibnamefont{Hawrylak}},
  \bibinfo{journal}{Phys. Rev. B} \textbf{\bibinfo{volume}{71}},
  \bibinfo{pages}{035316} (\bibinfo{year}{2005}).

\bibitem[{\citenamefont{McMurchie and Davidson}(1978)}]{mcmurchie1978one}
\bibinfo{author}{\bibfnamefont{L.~E.} \bibnamefont{McMurchie}}
  \bibnamefont{and} \bibinfo{author}{\bibfnamefont{E.~R.}
  \bibnamefont{Davidson}}, \bibinfo{journal}{J. Comput. Phys.}
  \textbf{\bibinfo{volume}{26}}, \bibinfo{pages}{218} (\bibinfo{year}{1978}).

\end{thebibliography}

\end{document}